\documentclass[aip,pop,reprint]{revtex4-1}
\usepackage{graphicx}
\usepackage{breqn}
\usepackage{geometry}
\usepackage{amssymb}
\usepackage{float}
\usepackage{eqnarray}
\usepackage{tabularx}
\usepackage{lipsum}

\newcommand{\leapp}{\stackrel{<}{{}_\sim}}

\usepackage[breaklinks=true,colorlinks=true,linkcolor=blue,urlcolor=blue,citecolor=blue]{hyperref}

\begin{document}

\title{Non-Maxwellian fast particle effects in gyrokinetic GENE simulations}

\author{A.~Di Siena}
\email{alessandro.di.siena@ipp.mpg.de}
\affiliation{Max Planck Institute for Plasma Physics  Boltzmannstr~2  85748~Garching Germany}
\author{T.~G\"orler}
\affiliation{Max Planck Institute for Plasma Physics  Boltzmannstr~2  85748~Garching Germany}
\author{H.~Doerk}
\affiliation{Max Planck Institute for Plasma Physics  Boltzmannstr~2  85748~Garching Germany}
\author{R.~Bilato}
\affiliation{Max Planck Institute for Plasma Physics  Boltzmannstr~2  85748~Garching Germany}
\author{J.~Citrin}
\affiliation{DIFFER Dutch Institute for Fundamental Energy Research De Zaale 20 5612 AJ Eindhoven The Netherlands}
\author{T.~Johnson}
\affiliation{VR Association EES KTH Stockholm Sweden}
\author{M.~Schneider}
\affiliation{CEA IRFM F13108 Saint Paul Lez Durance France}
\author{E.~Poli}
\affiliation{Max Planck Institute for Plasma Physics  Boltzmannstr~2  85748~Garching Germany}
\author{JET Contributors}
\affiliation{See the author list of  X Litaudon et al 2017 Nucl Fusion 57 102001}

\begin{abstract}
Fast ions have recently been found to significantly impact and partially suppress plasma turbulence both in experimental and numerical studies in a number of scenarios. Understanding the underlying physics and identifying the range of their beneficial effect is an essential task for future fusion reactors, where highly energetic ions are generated through fusion reactions and external heating schemes. However, in many of the gyrokinetic codes fast ions are, for simplicity, treated as equivalent-Maxwellian-distributed particle species, although it is well known that to rigorously model highly non-thermalised particles, a non-Maxwellian background distribution function is needed. To study the impact of this assumption, the gyrokinetic code GENE has recently been extended to support arbitrary background distribution functions which might be either analytic, e.g. slowing down and bi-Maxwellian, or obtained from numerical fast ion models. A particular JET plasma with strong fast-ion related turbulence suppression is revised with these new code capabilities both with linear and nonlinear gyrokinetic simulations. It appears that the fast ion stabilization tends to be less strong but still substantial with more realistic distributions, and this improves the quantitative power balance agreement with experiments.
\end{abstract}

\date{\today}
\maketitle

\section{Introduction}
A major factor limiting the performance of a fusion reactor is plasma turbulence. It is inevitably driven by steep temperature and density profiles and is one of the main reasons for the energy confinement degradation of nowadays tokamaks. In particular, the ion-temperature-gradient (ITG) instability has been identified as an important driver of microturbulence \cite{Romanelli_ITG}. Any mechanism able to reduce its development is extremely valuable and can lead to an increase of the energy confinement time. Among the different stabilising effects on the ITG microinstability, the presence of fast ions, generated through fusion reactions and/or external heating schemes, has recently been found to have a significant impact on plasma turbulence. Several studies have indeed shown that fast ions can passively dilute the main ion species \cite{Tardini_NF2007,Holland}, increase geometric stabilisation, i.e.~Shafranov shift stabilization \cite{Bourdelle}, reduce the thermal ITG drive through a wave fast ion resonance \cite{Di_Siena}, and finally actively stabilise linear growth rates and nonlinear fluxes through an electromagnetic stabilization related to fast ion suprathermal pressure gradients \cite{Romanelli,Citrin}. Other works have instead tackled the opposite issue, namely in which degree turbulence affects the fast ion background distribution function and the associated pressure profiles \cite{Wilkie_PPCF, Wilkie_PoP}. Thanks to these works significant progress in the understanding of fast ion affected turbulence and vice versa has already been made and a good agreement between numerical and experimental results is often achieved. However, in some of the most prominent studies where fast ions were found to be crucial to obtain realistic heat flux levels, the turbulence suppression appeared to be overestimated and power balance was, e.g., only reached with an increased main ion pressure gradient profile. In these works, an equivalent Maxwellian distribution function was employed for the highly non-thermalised fast ion species. Here, for the first time we address the impact of fast ions by using realistic distribution functions in the gyrokinetic code GENE \cite{GENE}. The latter has been modified recently to account for completely arbitrary background distribution functions, which might be either analytic, e.g. slowing down and Bi-Maxwellian, or numerical, e.g. extracted from specialized beam modelling codes like NEMO/SPOT \cite{NEMO} (the Neutral Beam Injected (NBI) particles) and SELFO \cite{SELFO} or TORIC/SSFPQL \cite{TORIC1, TORIC2} (for Ion Cyclotron Resonance Heated (ICRH) ions). The associated modifications in the underlying equations and in the source code will be discussed in the following before this new code version will be applied to one of the aforementioned scenarios with substantial fast-ion related turbulence suppression.
In detail, this paper is organized as follows. In Sec.~\ref{GENE_the} the basic gyrokinetic equations are discussed, for the full electromagnetic case, without any assumption on the background distribution function. The Vlasov equation, the moments of the distribution function and the Maxwell equations are self-consistently treated on the GENE coordinate grid. The limit of validity of the above derivation is discussed. In Sec.~\ref{exp_desc} an introduction of a JET L-mode discharge studied in this paper is presented and the non-Maxwellian distribution functions used in the GENE numerical simulations are defined in Sec.~\ref{equilibrium}. A linear and nonlinear analysis with the more realistic distribution functions for the fast ion species is respectively shown in Sec.~\ref{linear} and Sec.~\ref{nonlinear} and finally in Sec.~\ref{concl} general conclusions are drawn.

\section{Non-Maxwellian gyrokinetic equations} \label{GENE_the}

All the simulations presented in this work have been performed with the gyrokinetic code GENE, which solves numerically the Vlasov-Maxwell system of equations on a five dimensional grid for each time step. GENE can either be operated in the local flux tube approximation \cite{GENE}, in a radially global torus geometry \cite{Tobias} or as a flux-surface code \cite{Xanthopoulos}. Furthermore, full electromagnetic effects, realistic collision operators \cite{Tobias} and experimental geometries can be included. In the following section the basic gyrokinetic equations are re-derived in the full electromagnetic case without any assumption on the shape of the background distribution function. This general derivation allows a very flexible treatment of non-thermalised fast ion species, able to capture asymmetries and anisotropies of the background distribution function which might arise from the different heating schemes of a tokamak reactor.

\subsection{Vlasov equation}

The Vlasov equation determines the time evolution of the distribution function of each plasma species and in the gyro-center coordinate system $(\mathbf{X},v_\shortparallel,\mu)$ can be written as follows \cite{Brizard}
\begin{dmath}
\frac{\partial F}{\partial t}+\frac{d\mathbf{X}}{dt}\cdot\nabla F+\frac{dv_{\shortparallel}}{dt}\frac{\partial F}{\partial v_{\shortparallel}}+\frac{d\mu}{dt}\frac{\partial F}{\partial\mu}=0.
\label{eq:Vlasov0}
\end{dmath}
Here, $\mathbf{X}$ represents the centre of gyration, $v_\shortparallel$ the velocity along the magnetic field line and $\mu$ the magnetic moment. Expliciting the time derivatives of the coordinates\cite{Tobias}, Eq.~\ref{eq:Vlasov0} can be written as
\begin{dmath}
\frac{\partial F}{\partial t}+\left[v_{\shortparallel}\hat{b}_{0}+\left(\vec{v}_{E}+\vec{v}_{\nabla B}+\vec{v}_{c}\right)\right]\cdot\left\{ \vec{\nabla}F-\left[q\vec{\nabla}\bar{\phi}_{1}+\frac{q}{c}\hat{b}_{0}\dot{\bar{A}}_{1,\shortparallel}+\mu\vec{\nabla}\left(B_{0}+\bar{B}_{1,\shortparallel}\right)\right]\frac{1}{mv_{\shortparallel}}\frac{\partial F}{\partial v_{\shortparallel}}\right\} =0.
\label{eq:Vlasov1}
\end{dmath}
Here, the curvature, $E\times B_0$ and $\nabla B_0$ drift velocities have been defined as $\vec{v}_{c}=\frac{v_{\shortparallel}^{2}}{\Omega}\left(\vec{\nabla}\times\hat{b}_{0}\right)_{\perp}$, $\vec{v}_{E}=\frac{c}{B_{0}^{2}}\left(\vec{B}_{0}\times\vec{\nabla}\bar{\xi}_{1}\right)$ and $\vec{v}_{\nabla B_{0}}=\frac{\mu}{qB_{0}^{2}}\left(\vec{B}_{0}\times\vec{\nabla}B_{0}\right)$. Furthermore, $\bar{\xi}_{1}$ denotes the modified potential $\bar{\xi}_{1}=\bar{\phi}_{1}-\frac{v_{\shortparallel}}{c}\bar{A}_{1,\shortparallel}+\frac{\mu}{q}\bar{B}_{1,\shortparallel}$; $\Omega=\frac{qB_{0}}{mc}$ and $\hat{b}_{0}=\frac{\vec{B}_{0}}{B_{0}}$. The overbar denotes gyroaveraged quantities, which in the local code approximation reduce to the mere multiplication of Bessel functions, i.e. $\bar{\phi}_{1}=J_{0}\left(\lambda\right)\phi_{1}$; $\bar{A}_{1,\shortparallel}=J_{0}\left(\lambda\right)A_{1,\shortparallel}$ and $\bar{B}_{1,\shortparallel}=I_{1}\left(\lambda\right)B_{1,\shortparallel}$; where $I_{1}\left(\lambda\right)=\frac{2}{\lambda}J_{1}\left(\lambda\right)$ and $\lambda=\frac{k_{\perp}}{\Omega}(\frac{2B_{0}\mu}{m})^{1/2}$. Furthermore, $B_0$ denotes the background magnetic field; $E$ the perturbed electric field defined as $\vec{E} = \vec{\nabla}\bar{\xi}_{1}$; q and m, respectively, the charge and the mass of the considered species; c the speed of light and $k_{\perp}$ the perpendicular wavenumber.

An often employed approach in gyrokinetics is the splitting of the distribution function of each species into a background component and in a small fluctuating part, i.e. $F = F_0 + F_1$ (so-called $\delta f$). While many derivations like the previous one for GENE rely on local Maxwellian distributions, here we relax such assumption on $F_0$. The gyrokinetic $\delta f$ ordering, i.e. $n_{1}/n_{0}\sim\epsilon\ll1$, allows to greatly simplify the numerical solution of Eq.~\ref{eq:Vlasov1}. It is indeed possible to separate the time scale of variation of the background to the one of the fluctuating quantities through the expansion parameter $\epsilon$. The zeroth order term of the Vlasov equation, which reads as
\begin{dmath}
\frac{\partial F_{0}}{\partial t}=\hat{b}_{0}\cdot\left(v_{\shortparallel}\vec{\nabla}F_{0}-\frac{\mu}{m}\vec{\nabla}B_{0}\frac{\partial F_{0}}{\partial v_{\shortparallel}}\right)
\label{eq:background}
\end{dmath}
is exactly zero for local Maxwellian background (defined in Eq.~\ref{eq:maxwellian}). The zeroth order quantities can hence be considered time independent on the turbulent time scale. For the case of an arbitrary background distribution functions, Eq.~\ref{eq:background} is not necessarily zero and the degree of violation of Eq.~\ref{eq:background} must be studied case by case. In section \ref{equilibrium} an accurate analysis on Eq.~\ref{eq:background} is done for the numerical distribution functions employed in this paper.
The turbulent evolution of the system is determined by the first order term of Eq.~\ref{eq:Vlasov1}. 
It is convenient, at this point, to introduce a field aligned coordinate system, defined through the metric coefficients $g^{ij}=\nabla u^{i}\cdot\nabla u^{j}$, with $u^{i}=(x,y,z)$, $x$ radial direction, $y$ binormal direction, $z$ toroidal direction. The strong anisotropy of plasma turbulence respect to the magnetic field, i.e. $k_{\perp}/k_{\shortparallel}<<1$, allows to greatly simplify the analytical derivation of the first order term of Eq.~\ref{eq:Vlasov1}, which becomes
\begin{dmath}
\frac{\partial g_{1}}{\partial t}+\frac{C}{JB_{0}}\left\{ v_{\shortparallel}\partial_{z}F_{1}-\left(\frac{q}{m}\partial_{z}\bar{\phi}_{1}\frac{\partial F_{0}}{\partial v_{\shortparallel}}+\frac{\mu}{m}\partial_{z}B_{0}\frac{\partial F_{1}}{\partial v_{\shortparallel}}+\frac{\mu}{m}\partial_{z}\bar{B}_{1,\shortparallel}\frac{\partial F_{0}}{\partial v_{\shortparallel}}\right)\right\} +\frac{c}{C}\left(\frac{g^{1i}g^{2j}-g^{2i}g^{1j}}{\gamma_{1}}\right)\left\{ \left[\partial_{i}\bar{\xi}_{1}+\frac{\mu}{q}\partial_{i}B_{0}+\frac{v_{\shortparallel}^{2}m}{q}\left(\frac{\partial_{i}B_{0}}{B_{0}}+\frac{\beta_{p}}{2}\frac{\partial_{i}p_{0}}{p_{0}}\right)\right]\cdot\left[\partial_{j}F_{0}+\partial_{j}F_{1}-\left(q\partial_{j}\bar{\phi}_{1}+\mu\partial_{j}B_{0}+\mu\partial_{j}\bar{B}_{1,\shortparallel}\right)\frac{1}{mv_{\shortparallel}}\frac{\partial F_{0}}{\partial v_{\shortparallel}}\right]\right\} =0.
\label{eq:Vlasov2}
\end{dmath}
A modified distribution function $g_{1}=F_{1}-\frac{q}{mc}\bar{A}_{1,\shortparallel}\frac{\partial F_{0}}{\partial v_{\shortparallel}}$ has been introduced and the following geometrical coefficients have been defined $\gamma_{1}=g^{11}g^{22}-g^{21}g^{12}$ and $C=B_{0}/\gamma_{1}^{1/2}$. Eq.~\ref{eq:Vlasov2} is solved in dimensionless units. With this aim, all the physical quantities have been split into a dimensionless value and a dimensional reference part. The reference values used for normalizing Eq.~\ref{eq:Vlasov2} are the elementary electron charge $e$, the main ion mass $m_{i}$ and temperature $T_{i}$, a reference magnetic field $B_{\rm{ref}}$ and a macroscopic length $L_{\rm{ref}}$. The normalized Vlasov equation for a completely general background distribution function can be written as follows
\begin{widetext}
\begin{dmath}
\frac{\partial g_{1}}{\partial t}=-\frac{C}{JB_{0}}v_{th}v_{\shortparallel}\left[\partial_{z}F_{1}-\frac{q}{2T_{0}v_{\shortparallel}}\partial_{z}\phi_{1}\frac{\partial F_{0}}{\partial v_{\shortparallel}}-\frac{\mu}{2v_{\shortparallel}}\partial_{z}B_{1,\shortparallel}\frac{\partial F_{0}}{\partial v_{\shortparallel}}\right]+\frac{C}{JB_{0}}v_{th}\frac{\mu}{2}\partial_{z}B_{0}\frac{\partial F_{1}}{\partial v_{\shortparallel}}+\frac{T_{0}}{q}\left(\frac{\mu B_{0}+2v_{\shortparallel}^{2}}{B_{0}}\right)\mathcal{K}_{x}\hat{\partial}_{x}F_{0}-\frac{T_{0}}{q}\left[\left(\frac{\mu B_{0}+2v_{\shortparallel}^{2}}{B_{0}}\right)\mathcal{K}_{y}-\frac{1}{C}\frac{v_{\shortparallel}^{2}\beta_{ref}}{B_{0}^{2}}\omega_{p}\right]\partial_{y}g_{1}+\left[\frac{1}{2v_{\shortparallel}}\frac{\partial F_{0}}{\partial v_{\shortparallel}}\left(\frac{\mu B_{0}+2v_{\shortparallel}^{2}}{B_{0}}\right)\mathcal{K}_{y}-\frac{1}{C}\frac{v_{\shortparallel}^{2}\beta_{ref}}{B_{0}^{2}}\omega_{p}\frac{1}{2v_{\shortparallel}}\frac{\partial F_{0}}{\partial v_{\shortparallel}}-\frac{1}{C}\hat{\partial}_{x}F_{0}\right]\partial_{y}\xi_{1}-\frac{T_{0}}{q}\left(\frac{\mu B_{0}+2v_{\shortparallel}^{2}}{B_{0}}\right)\mathcal{K}_{x}\partial_{x}g_{1}+\frac{1}{2v_{\shortparallel}}\frac{\partial F_{0}}{\partial v_{\shortparallel}}\left(\frac{\mu B_{0}+2v_{\shortparallel}^{2}}{B_{0}}\right)\mathcal{K}_{x}\partial_{x}\xi_{1}-\frac{1}{C}\left[\partial_{x}\xi_{1}\partial_{y}g_{1}-\partial_{y}\xi_{1}\partial_{x}g_{1}\right]
\label{eq:Vlasov3}
\end{dmath}
\end{widetext}
where the following geometrical coefficients $\mathcal{K}_{x}=-\frac{1}{C}\left(\partial_{y}B_{0}-\frac{\gamma_{3}}{\gamma_{1}}\partial_{z}B_{0}\right)$, $\mathcal{K}_{y}=\frac{1}{C}\left(\partial_{x}B_{0}-\frac{\gamma_{3}}{\gamma_{1}}\partial_{z}B_{0}\right)$; the normalized x-derivative $\hat{\partial}_x = -(\partial_x - \frac{\mu}{2v_{\shortparallel}}\partial_x B_0 \frac{\partial}{\partial v_\shortparallel})$; the normalized background pressure gradient $\omega_{p}=-L_{\rm{ref}}\frac{\partial_{x}p_{0}}{n_{\rm{ref}}T_{\rm{ref}}}$ and the reference thermal to magnetic pressure ratio $\beta_{\rm{ref}}=\frac{8\pi n_{\rm{ref}}T_{\rm{ref}}}{B_{\rm{ref}}^{2}}$  have been defined. If the equilibrium distribution function $F_{0}$ is a local Maxwellian, it can be shown that Eq.~\ref{eq:Vlasov3} reduces to the gyrokinetic equation known in literature \cite{Tobias,Brizard}.

\subsection{Velocity space moments}

In order to treat self-consistently the Vlasov-Maxwell system of coupled equations, the fluctuating component of the fields must be evaluated from the perturbed distribution function of each plasma species at every time step. For this reason, in the following section a general description of the moments of $F_1$, which enter in the calculation of the field components, is presented without making any assumptions on the background distribution function. The general $a$-$\rm{th}$ moment in $v_\shortparallel$ and $b$-$\rm{th}$ in $\mu$ (or, more precisely, in $v_\perp$) in the guiding centre coordinate system ($\mathbf{X},\theta,v_\shortparallel,\mu$) is defined as follows
\begin{widetext}
\begin{dmath}
M_{a,b}\left(\mathbf{x}\right)=2^{b/2}\left(\frac{B_{0}}{m}\right)^{b/2+1}\int\delta\left(\mathbf{X}+\mathbf{r}-\mathbf{x}\right)f^{gc}_{1}\left(\mathbf{X},\theta,v_{\shortparallel},\mu\right)v_{\shortparallel}^{a}\mu^{b/2}d^{3}Xdv_{\shortparallel}d\mu d\theta.
\label{eq:moment1}
\end{dmath}
\end{widetext}
Here, $f^{gc}_{1}\left(\mathbf{X},\theta,v_{\shortparallel},\mu\right)$ is the perturbed distribution function in the guiding centre coordinate system and $\delta$ is the Dirac-delta function. The space transformation used to link the particle coordinates to the guiding centre is $\mathbf{x} = \mathbf{X} + \mathbf{r}(\mathbf{X},\mu,\theta)$, where $\mathbf{r}$ denotes the gyroradius vector. Since, from Eq.~\ref{eq:Vlasov3} the time evolution of the perturbed distribution function is performed in the gyro centre coordinate system it is necessary to define an operator $T^{*}$ which transforms $F_1$ from the gyro centre to the guiding centre coordinate system. $T^{*}$ is defined up to the first order in the gyrokinetic expansion \cite{Brizard,Dannert} as follows
\begin{dmath}
f^{gc}_{1}\left(\mathbf{X},\theta,v_{\shortparallel},\mu\right)=T^{*}F_{1}\left(\mathbf{X},v_{\shortparallel},\mu\right)
\end{dmath}
\begin{dmath}
=F_{1}+\frac{1}{B_{0}}\left\{ \left[\Omega\frac{\partial F_{0}}{\partial v_{\shortparallel}}-\frac{q}{c}v_{\shortparallel}\frac{\partial F_{0}}{\partial\mu}\right]\left(A_{1,\shortparallel}\left(\mathbf{X}+\mathbf{r}\right)-\bar{A}_{1,\shortparallel}\left(\mathbf{X}\right)\right)+\left[q\left(\phi_{1}\left(\mathbf{X}+\mathbf{r}\right)-\bar{\phi}_{1}\left(\mathbf{X}\right)\right)-\mu\bar{B}_{1,\shortparallel}\right]\frac{\partial F_{0}}{\partial\mu}\right\}.
\label{eq:pullback}
\end{dmath}
By performing the integrals over $\theta$ and $\mathbf{X}$ and using the previously defined operator, the generic moment of the gyro centre distribution function reduces to
\begin{widetext}
\begin{dmath}
M_{a,b}\left(\mathbf{x}\right)=\pi\left(\frac{2B_{0}}{m}\right)^{b/2+1}\int\left\{ \left\langle F_{1}\left(\mathbf{x}-\mathbf{r}\right)\right\rangle +\left(\frac{\Omega}{B_{0}}\frac{\partial F_{0}}{\partial v_{\shortparallel}}-\frac{q}{cB_{0}}v_{\shortparallel}\frac{\partial F_{0}}{\partial\mu}\right)\left(A_{1,\shortparallel}\left(\mathbf{x}\right)-\left\langle \bar{A}_{1,\shortparallel}\left(\mathbf{x}-\vec{r}\right)\right\rangle \right)+\left[\frac{q}{B_{0}}\left(\phi_{1}\left(\mathbf{x}\right)-\left\langle \bar{\phi}_{1}\left(\mathbf{x}-\mathbf{r}\right)\right\rangle \right)-\frac{\mu}{B_{0}}\left\langle \bar{B}_{1,\shortparallel}\left(\mathbf{x}-\mathbf{r}\right)\right\rangle \right]\frac{\partial F_{0}}{\partial\mu}\right\} v_{\shortparallel}^{a}\mu^{b/2}dv_{\shortparallel}d\mu,
\label{eq:moment3}
\end{dmath}
\end{widetext}
where $\left\langle ...\right\rangle =\frac{1}{2\pi}\int...d\theta$. In the specific case of a Maxwellian background, Eq.~\ref{eq:moment3} can be greatly simplified, i.e.~the term that multiplies the vector potential is exactly zero. As it will be shown in the next section, the latter simplification leads to a decoupling between the Poisson and $B_{1,\shortparallel}$ equations and the parallel component of the Ampere's law. This is not necessarily the case for non-Maxwellian distribution function.

\subsection{Field equations}

The Poisson equation and the Ampere's law for both the parallel and perpendicular component of the electromagnetic potential can be written in terms of the $M_{0,0}$, $M_{1,0}$ and $M_{0,1}$ moments of the perturbed distribution function $F_1$ as follows
\begin{dmath}
\nabla_{\perp}^{2}\phi_{1}\left(\mathbf{x}\right)=-4\pi\sum_{j}q_{j}n_{1,j}\left(\mathbf{x}\right)=-4\pi\sum_{j}q_{j}M_{0,0,j}\left(\mathbf{x}\right),
\label{eq:Poisson1}
\end{dmath}

\begin{dmath}
-\nabla_{\perp}^{2}A_{1,\shortparallel}\left(\mathbf{x}\right)=\frac{4\pi}{c}\sum_{j}j_{\shortparallel,1,j}\left(\mathbf{x}\right)=\frac{4\pi}{c}q_{j}M_{1,0,j}\left(\mathbf{x}\right),
\label{eq:Ampere1}
\end{dmath}

\begin{dmath}
\hat{e}_{1}\partial_{y}B_{1,\shortparallel}\left(\mathbf{x}\right)+\hat{e}_{2}\partial_{x}B_{1,\shortparallel}\left(\mathbf{x}\right)=\frac{4\pi}{c}\sum_{j}\vec{j}_{1,\perp,j}\left(\mathbf{x}\right)=\frac{4\pi}{c}\sum_{j}q_{j}\hat{c}(\theta)M_{0,1,j}\left(\mathbf{x}\right).
\label{eq:Bpara1}
\end{dmath}
The field equations have been written in the particle coordinate system $\left(\hat{e}_{1},\hat{e}_{2},\hat{b}_{0}\right)$, where $\hat{c}(\theta)$ is the unit vector in the perpendicular plane, $\hat{c}(\theta) = -\sin\theta \hat{e}_1+\cos \theta\hat{e}_2$. From Eq.~\ref{eq:moment3} it is possible to reformulate the field equations in terms of the perturbed distribution function $F_1$ as it is done in Eq.~\ref{eq:Poisson2},~\ref{eq:Ampere2},~\ref{eq:Bpara2}. For the sake of simplicity, in the following equations the sum over all species is omitted.

\begin{dmath}
P\phi_{1}\left(\mathbf{x}\right)+\mathcal{F}A_{1,\shortparallel}\left(\mathbf{x}\right)+\mathcal{T}B_{1,\shortparallel}\left(\mathbf{x}\right)=q\pi n_{0}B_{0}\int J_{0}g_{1}\left(\mathbf{x}\right)dv_{\shortparallel}d\mu
\label{eq:Poisson2}
\end{dmath}

\begin{dmath}
\mathcal{L}\phi_{1}\left(\mathbf{x}\right)+\mathcal{H}A_{1,\shortparallel}\left(\mathbf{x}\right)+\mathcal{K}B_{1,\shortparallel}\left(\mathbf{x}\right)=qn_{0}\pi\beta_{ref}\frac{B_{0}v_{th}}{2}\int v_{\shortparallel}J_{0}g_{1}\left(\mathbf{x}\right)dv_{\shortparallel}d\mu
\label{eq:Ampere2}
\end{dmath}

\begin{dmath}
\mathcal{R}\phi_{1}\left(\mathbf{x}\right)+\mathcal{W}A_{1,\shortparallel}\left(\mathbf{x}\right)+\mathcal{Q}B_{1,\shortparallel}\left(\mathbf{x}\right)=B_{0}^{\frac{3}{2}}\frac{q\pi n_{0}v_{th}}{2k_{\perp}}\beta_{ref}\int\sqrt{\mu}J_{1}g_{1}\left(\mathbf{x}\right)dv_{\shortparallel}d\mu
\label{eq:Bpara2}
\end{dmath}
The following operators have been defined

\begin{dmath}
P=k_{\perp}^{2}\lambda_{De}^{2}-\frac{\pi q^{2}n_{0}}{T_{0}}\int\left(1-J_{0}^{2}\right)\frac{\partial F_{0}}{\partial\mu}dv_{\shortparallel}d\mu
\end{dmath}

\begin{dmath}
\mathcal{F}=\frac{2\pi q^{2}n_{0}}{mv_{th}}\int\left[\left(1-J_{0}^{2}\right)v_{\shortparallel}\frac{\partial F_{0}}{\partial\mu}-\frac{B_{0}}{2}\frac{\partial F_{0}}{\partial v_{\shortparallel}}\right]dv_{\shortparallel}d\mu
\end{dmath}

\begin{dmath}
\mathcal{T}=\pi qn_{0}\int\mu J_{0}I_{1}\frac{\partial F_{0}}{\partial\mu}dv_{\shortparallel}d\mu
\end{dmath}

\begin{dmath}
\mathcal{H}=k_{\perp}^{2}-\frac{q^{2}n_{0}\pi\beta_{ref}}{m}\int\left[B_{0}\frac{v_{\shortparallel}}{2}\frac{\partial F_{0}}{\partial v_{\shortparallel}}-v_{\shortparallel}^{2}\frac{\partial F_{0}}{\partial\mu}\left(1-J_{0}^{2}\right)\right]dv_{\shortparallel}d\mu
\end{dmath}

\begin{dmath}
\mathcal{L}=\frac{q^{2}n_{0}\pi\beta_{ref}}{mv_{th}}\int\frac{\partial F_{0}}{\partial\mu}\left(1-J_{0}^{2}\right)v_{\shortparallel}dv_{\shortparallel}d\mu
\end{dmath}

\begin{dmath}
\mathcal{K}=\frac{v_{th}}{2}\int J_{0}I_{1}v_{\shortparallel}\mu\frac{\partial F_{0}}{\partial\mu}dv_{\shortparallel}d\mu
\end{dmath}

\begin{dmath}
\mathcal{Q}=-1+\frac{\pi q^{2}n_{0}B_{0}}{mk_{\perp}^{2}}\beta_{ref}\int\mu\frac{\partial F_{0}}{\partial\mu}J_{1}^{2}dv_{\shortparallel}d\mu
\end{dmath}

\begin{dmath}
\mathcal{W}=-B_{0}^{\frac{1}{2}}\frac{q^{2}n_{0}}{mk_{\perp}}\beta_{ref}\int\sqrt{\mu}v_{\shortparallel}J_{1}J_{0}\frac{\partial F_{0}}{\partial\mu}dv_{\shortparallel}d\mu
\end{dmath}

\begin{dmath}
\mathcal{R}=B_{0}^{\frac{1}{2}}\frac{\pi q^{2}n_{0}}{mk_{\perp}v_{th}}\beta_{ref}\int\sqrt{\mu}\frac{\partial F_{0}}{\partial\mu}J_{1}J_{0}dv_{\shortparallel}d\mu
\end{dmath}
For a completely general background distribution function each component of the fields is coupled to the others. This system decouples for the $A_{1,\shortparallel}$ component if a Maxwellian distribution functions is chosen, since $\mathcal{F} = \mathcal{W} = \mathcal{L} = \mathcal{K} = 0$. 

\section{Application of realistic fast particle background distributions} \label{exp_desc}

\begin{figure*}
\begin{center}
\includegraphics[scale=0.28]{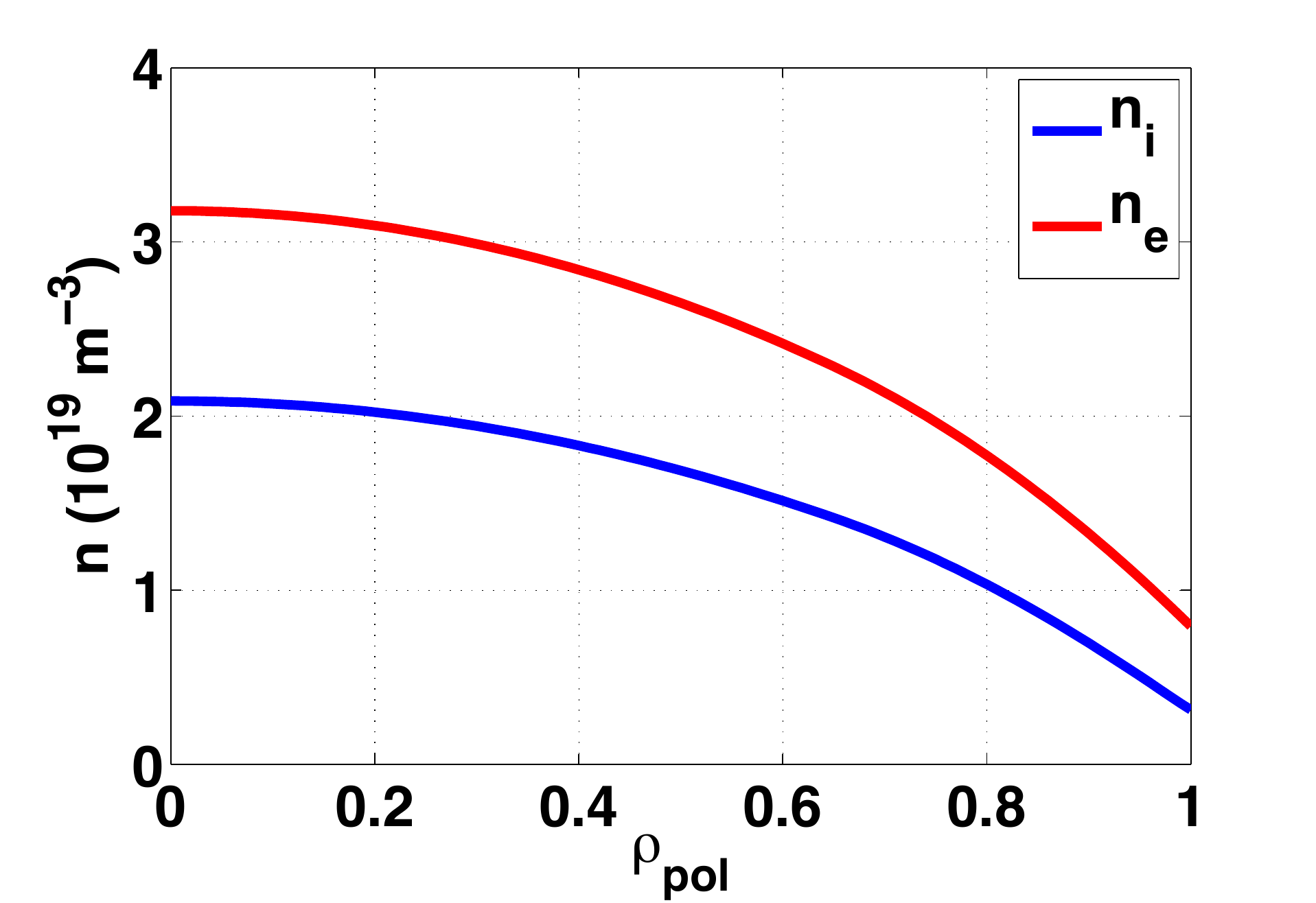}\includegraphics[scale=0.28]{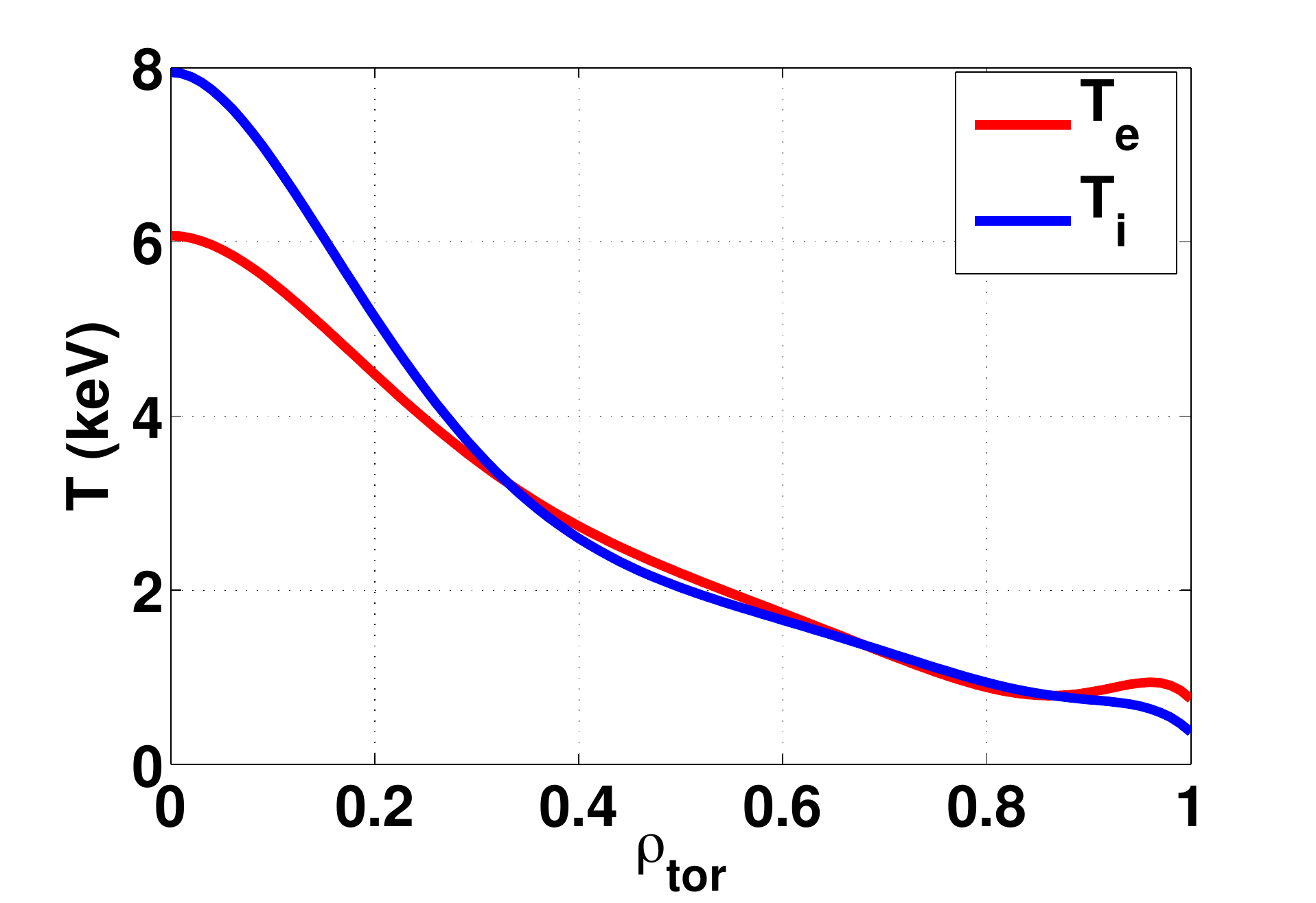}
\par\end{center}
\caption{Radial profiles of main ions (blue line) and electron (red line) a) temperature and b) density for the discharge 73224.}
\label{fig:profiles}
\end{figure*}

Taking advantage of these new capabilities of the gyrokinetic code GENE, experimental discharges associated to significant fast ion stabilisation can now be studied with the more realistic modelling tools for the energetic ion population introduced in Sec.~\ref{GENE_the}. The newly implemented terms have been benchmarked with the gyrokinetic codes GKW and GS2 for simplified geometry and in the electrostatic limit in Ref.~\onlinecite{Varenna} and in the work at hand a realistic scenario is extensively studied. The JET C-wall L-mode plasma 73224 has been selected and re-analysed with the more realistic non-Maxwellian distribution functions. The experiment was performed with vacuum toroidal magnetic field $B_T \approx 3.3 T$, plasma current $Ip \approx 2 MA$ and with $q_{95} \approx 6$. The heating power consists of $3.5 MW$ of ICRH in $(3He) − D$ minority scheme and of $1.5 MW$ of NBI. Furthermore, the ICRH power was deposited on-axis. The plasma was composed of bulk thermal Deuterium, electron and Carbon impurities and of fast NBI Deuterium and ICRH $^{3}$He. An accurate description of this discharge can be found in Ref.~\onlinecite{Mantica_PRL2009,Mantica_PRL2011,Citrin_PRL2013}. Experimental geometry, collisions (Landau-Boltzmann operator), electromagnetic fluctuations and kinetic electrons are included. The magnetic geometry and the nominal plasma parameters are summarised in table ~\ref{table:parameters} and the radial thermal density and temperature profiles, reconstructed by CRONOS simulations, are shown in Fig.~\ref{fig:profiles}. The analysis of this discharge is performed in the local flux tube approximation at a radial position of $\rho_{\rm{tor}} = 0.33$, i.e. where a significant fast ion turbulence suppression is observed. The local approach is justified by low values of the ion Larmor radius normalized to the tokamak minor radius, i.e. $\rho_i / a$, with $\rho_i = (T_i/m_i)^{1/2} / \Omega$, namely $\rho^* = 1/450$ for thermal ions and $\rho_{fast,D}^* = 1/150$; $\rho_{^3He}^* = 1/200$, respectively, for fast deuterium and helium.
\begin{table*}
\caption{\label{table:parameters}Parameters at $\rho_{\rm{tor}} = 0.33$ for the JET L-mode discharge 73224 according to Ref.~\cite{Citrin_PRL2013,Bravenec_PPCF2016}. $T$ represents the temperature normalized to the electron one, $R/L_{T,n}$ the normalized logarithmic temperature and density gradients and $\nu^{*}$ the electron-ion collision frequency normalized to the trapped electron bounce frequency.}
\begin{ruledtabular}
\begin{tabular}{llllllll}
  R & $\hat{s}$ & q & $T_{e}/T_{i}$ & $R/L_{T_{i}}$ & $R/L_{T_{e}}$ & $R/L_{n_{e}}$ & $\nu^{*}$\\
  \hline
  3.1& 0.5 & 1.7 & 1.0 & 9.3 & 6.8 & 1.3 & 0.038 \\
  \hline
  $n_{fD}$ & $n_{^{3}He}$ & $T_{fD}$ & $T_{^{3}He}$ & $R/L_{T_{fD}}$ & $R/L_{T_{^{3}He}}$ & $R/L_{nfD}$ & $R/L_{n_{^{3}He}}$ \\
  \hline
  0.06 & 0.07 & 9.8 & 6.9 & 3.2 & 23.1 & 14.8 & 1.6\\
\end{tabular}
\end{ruledtabular}
\end{table*}

\subsection{Equilibrium distribution functions} \label{equilibrium}

\begin{figure*}
\begin{center}
\includegraphics[scale=0.27]{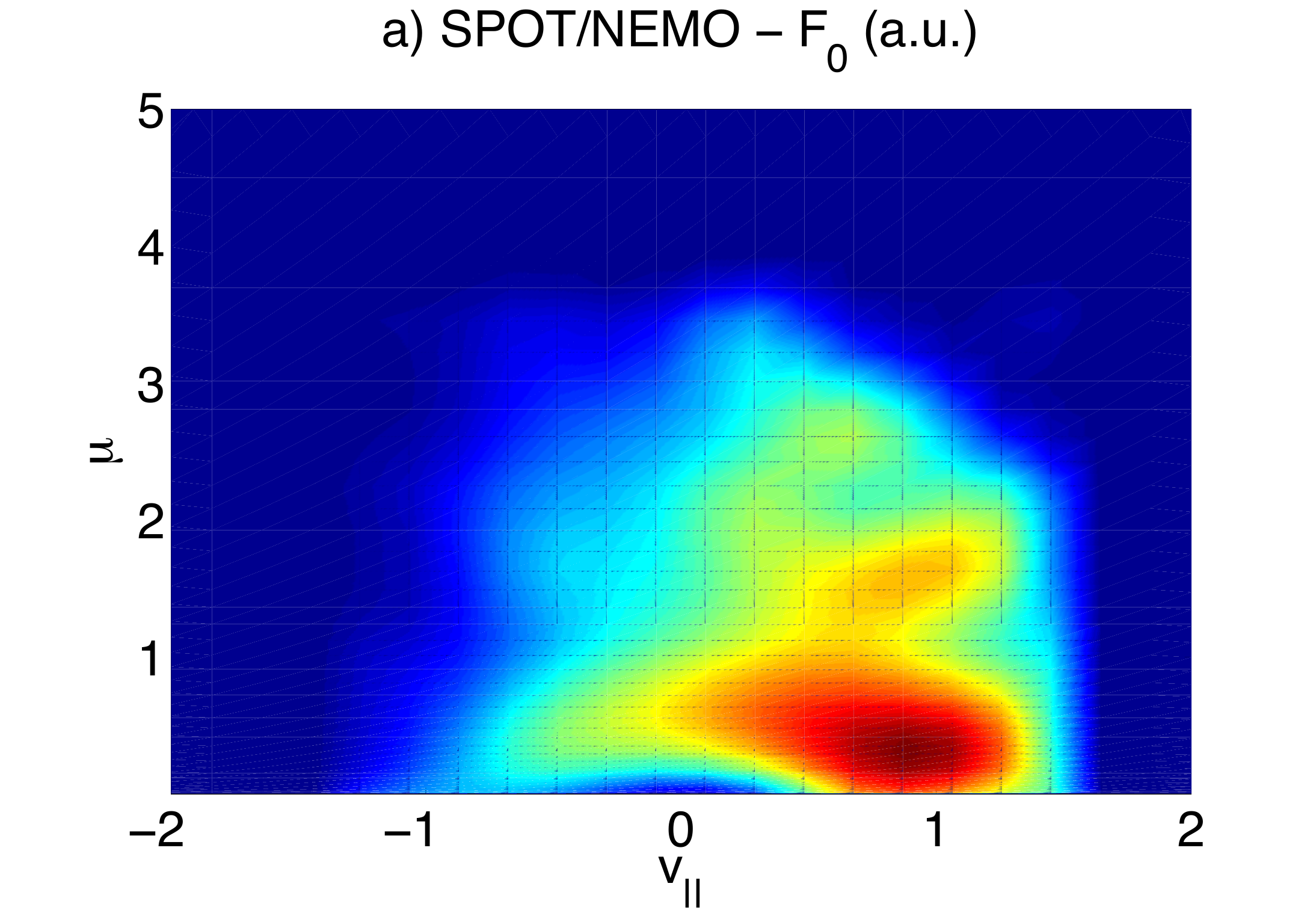}\includegraphics[scale=0.27]{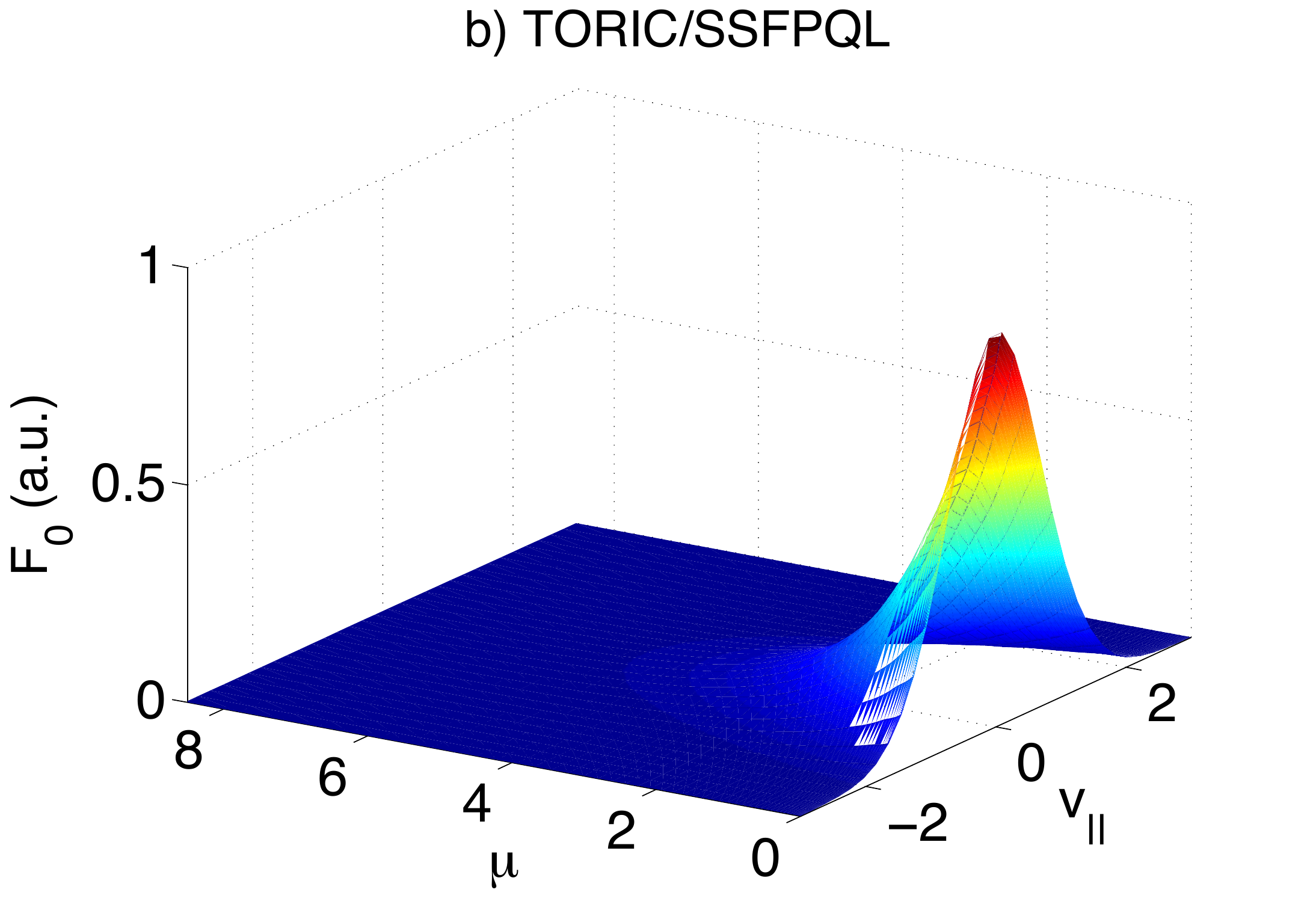}
\par\end{center}
\caption{$\theta$-integrated a) SPOT/NEMO and b) TORIC/SSFPQL numerical distribution functions on the $(v_\shortparallel,\mu)$ velocity grid.}
\label{fig:SPOT}
\end{figure*}

As mentioned in section \ref{GENE_the}, a $\delta f$ approach is employed for solving the gyrokinetic system of equations where the distribution function of each species is split into a time independent background component and a small fluctuating part. For all the thermal species, the background $F_0$ is assumed to be the local Maxwellian distribution function as defined as follows
\begin{equation}
F_{0,M}=\frac{n_{0}}{\pi^{3/2}v_{th}^{3}}\exp\left(\frac{-mv_{\shortparallel}^{2}/2-\mu B_{0}}{T_{0}}\right).
\label{eq:maxwellian}
\end{equation}
Here, $m$ is the particle mass, $T_{0}$ the equilibrium temperature, $n_{0}$ the particle density, $v_{th}=(2T_{0}/m)^{1/2}$ the thermal velocity and $B_0$ the equilibrium magnetic field. 
For the case of energetic ions the more flexible $F_0$ setup presented in Sec.~\ref{GENE_the} has been implemented in the code. GENE is able to support a large variety of different background distribution functions which can be either analytical or numerical. In so doing, it is possible to capture asymmetries and anisotropies in the distribution function arising from the different heating schemes, e.g. ICRH and NBI. In particular, here, different backgrounds are used for each fast particle species. 
For the case of NBI fast Deuterium a numerical distribution function has been extracted from SPOT/NEMO simulations with 4191 test particles and has been interpolated on the GENE coordinate grid. 
In Fig.~\ref{fig:SPOT}a) the numerical SPOT/NEMO distribution function is shown on the GENE $v_{\shortparallel}-\mu$ grid. A velocity space structure similar to a slowing down distribution can be identified with a cut-off velocity $v_{\shortparallel,c} \sim 1.5$. Furthermore, a strong velocity anisotropy between co-passing and counter-passing fast particles is observed.
In the next paragraph a linear analysis is performed studying the impact of the different backgrounds on the linear observables, i.e. growth rates and frequencies. Regarding the NBI fast deuterium, the results obtained with the SPOT/NEMO distribution function are compared to the ones obtained with the analytic slowing-down function derived in Ref.~\onlinecite{Gaffey} and, e.g., used in Ref.~\onlinecite{Angioni} for modelling fusion born alpha particles. The latter is a solution of the Fokker-Planck equation with an isotropic delta-function particle source and is defined as follows
\begin{equation}
F_{0,s}=\frac{3n_{0}}{4\pi\log\left(1+\frac{v_{\alpha}^{3}}{v_{c}^{3}}\right)\left[v_{c}^{3}+v^{3}\right]}\Theta\left(v_{\alpha}-v\right).
\label{eq:slowd}
\end{equation}
Here, the birth velocity is defined through the birth energy $E_{\alpha}$ in the following way $v_{\alpha}=(2E_{\alpha}/m_{\alpha})^{1/2}$ , while $v_{c}=v_{th,e}\left(\frac{3\sqrt{\pi}m_{e}}{4}\sum_{\rm main\:ions}\frac{n_{i}z_{i}^{2}}{n_{e}m_{i}}\right)^{1/3}$ represents the critical slowing down velocity. Furthermore,  $\Theta$ is the Heaviside step function.

For the case of the ICRH $^{3}$He, numerical distribution functions extracted from TORIC/SSFPQL and SELFO/LION+FIDO are used both in the linear and turbulence analysis presented in this work. Interface routines between these different codes and GENE have been implemented. The SPOT/NEMO and SELFO/LION+FIDO numerical distribution functions, here employed, had already been used in Ref.~\onlinecite{Citrin} to calculate the fast ion profiles for the equivalent Maxwellian distribution function respectively for the NBI and ICRH-driven fast ions.
In Fig.~\ref{fig:SPOT}b) the phase space structure of $^{3}$He distribution functions extracted from TORIC/SSFPQL is shown on the GENE coordinate grid. No significant difference with the SELFO/LION+FIDO background is observed.
As for the NBI fast Deuterium, a first order analytical approximation is applied for the ICRH $^{3}$He. To account for anisotropies in velocity arising from the ICRH heating scheme, a bi-Maxwellian distribution
\begin{equation}
F_{0,aM}=\frac{n_{0}}{\pi^{3/2}v_{th,\shortparallel}v_{th,\perp}^{2}}{ \exp(-v_{\shortparallel}^{2}/v_{th,\shortparallel}^{2}-\frac{\mu B_{0}}{T_{\perp}})}. 
\label{eq:bimax}
\end{equation}
is used through all the rest of this work.
Here, $T_{\shortparallel}$ and $T_{\perp}$ are respectively the parallel and perpendicular temperatures. The $T_\perp / T_\shortparallel = 2.2$ and $L_{T_\shortparallel} / L_{T_\perp} = 3$ anisotropies have been extracted from SELFO/LION+FIDO simulations and are consistent with the ones evaluated with TORIC/SSFPQL. Furthermore, the fast particle temperatures have been defined as the second order moment of the numerical distribution functions \cite{Estrada,Angioni}, i.e. NEMO/SPOT for the NBI fast deuterium and TORIC/SSFPQL and SELFO/LION+FIDO for the ICRH $^3$He, namely
\begin{equation}
T = \frac{\int v^{2}F_{\rm{0,numerical}}d^3v}{\int F_{\rm{0,numerical}}d^3v}.
\label{eq:press}
\end{equation}
\begin{figure}
\begin{center}
\includegraphics[scale=0.30]{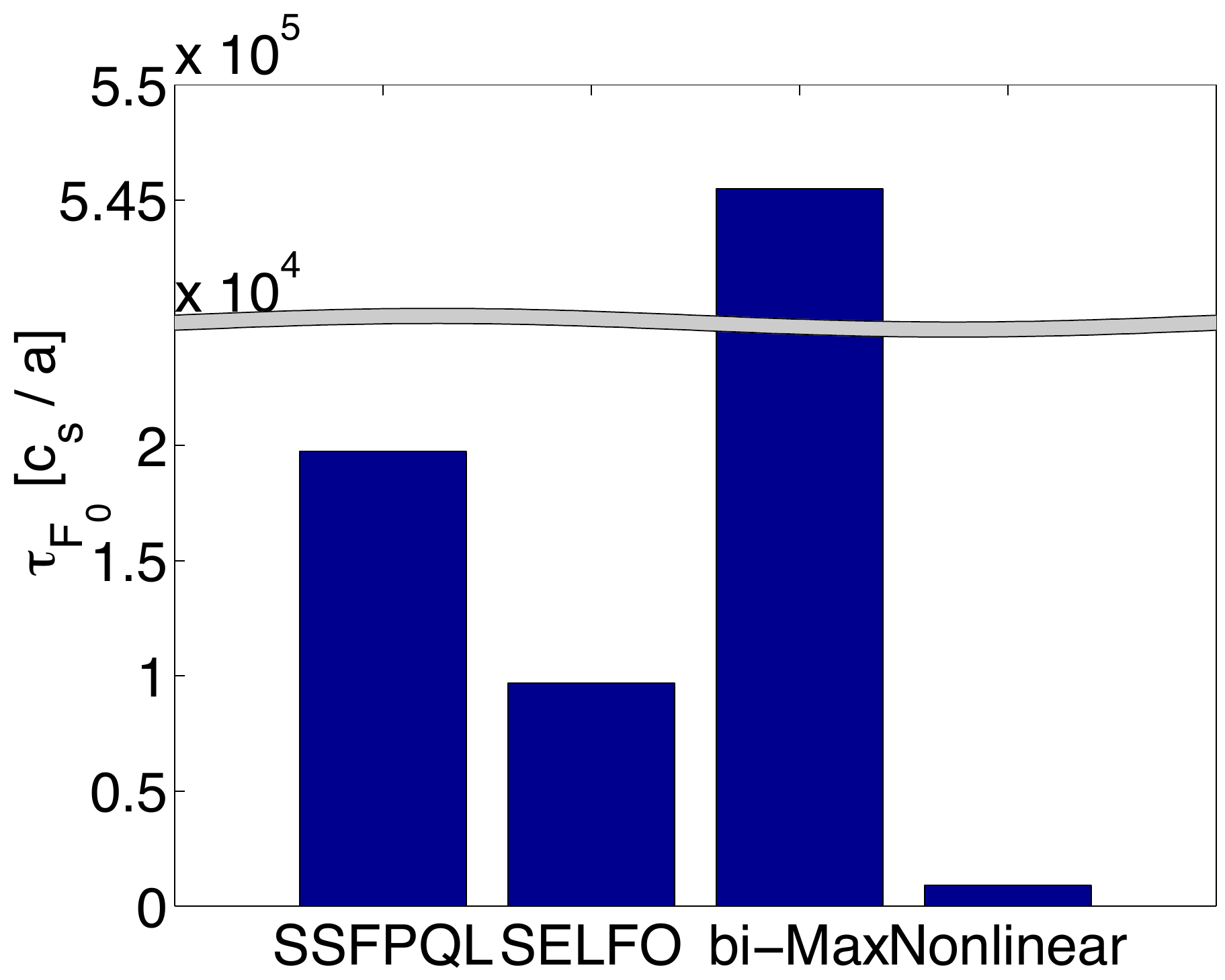}
\par\end{center}
\caption{Comparison between the time scale of variation $\tau_{F_0}$ of the backgrounds employed in the turbulence analysis of Sec.~\ref{nonlinear} and the average time of GENE nonlinear simulations in units of $c_s / a$.}
\label{fig:violation}
\end{figure}
One of the major constraints on the non-Maxwellian backgrounds is set by Eq.~\ref{eq:background}. The time scale of background distribution variations $\tau_{F_0}$ described by the zeroth order Vlasov equation should always be well separated from the turbulent time scale. While this can be easily shown to be the case for local Maxwellians and slowing down backgrounds, other distribution functions like the numerical and bi-Maxwellian ones require a more detailed study of Eq.~\ref{eq:background}. Corresponding results are shown in Fig.~\ref{fig:violation} and demonstrate that the average time - normalized to $c_s / a$ - required in the GENE nonlinear simulations to reach a saturated turbulence state is several order of magnitudes smaller than $\tau_{F_0}$. The background distributions can thus be considered constant in time.

\subsection{Linear growth rate analysis} \label{linear}

\begin{figure*}
\begin{center}
\includegraphics[scale=0.25]{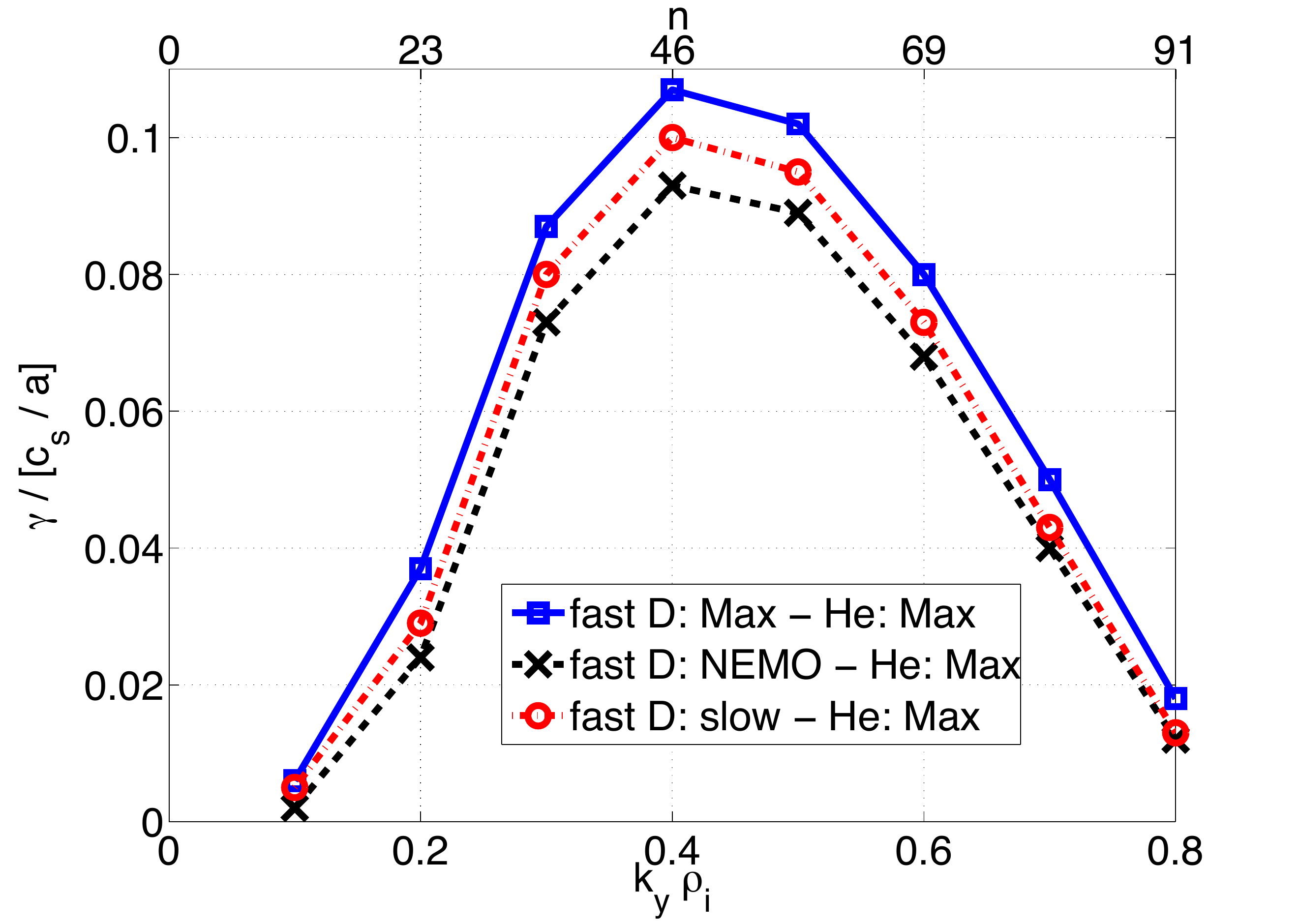}\includegraphics[scale=0.25]{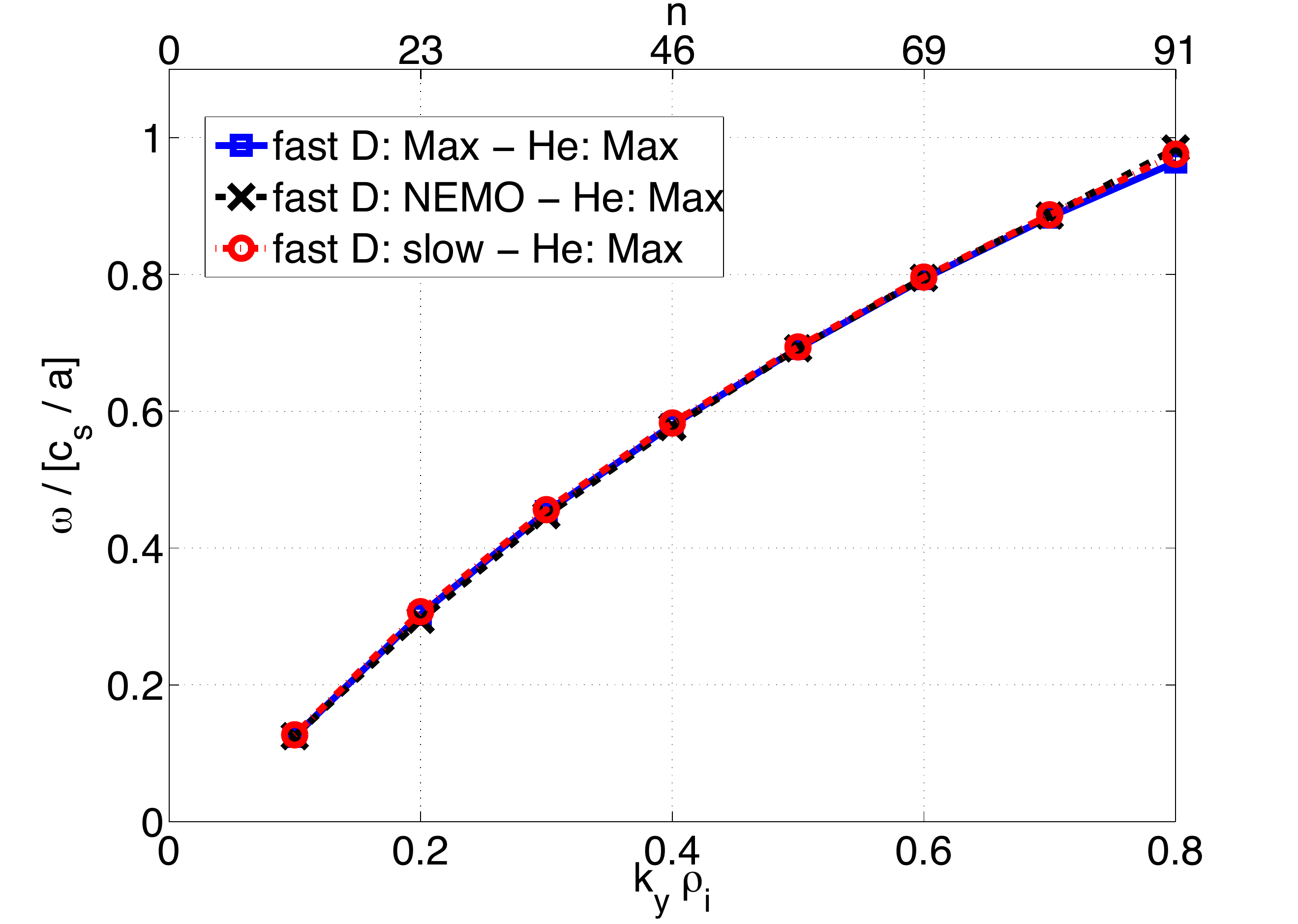}
\par\end{center}
\caption{GENE calculation of the linear growth rates (a) and frequency (b) for different $k_y \rho_i$ and toroidal mode numbers $n$ for different distribution functions for the fast Deuterium.}
\label{fig:nbi}
\end{figure*}

This section adresses the impact of the more realistic distribution functions on the ITG microturbulence. Although a true comparison with experiments can only be made with fully nonlinear simulations (see next session), it is still possible to extract valuable information about the expected nonlinear sensitivity of the ITG dominated physics on the different fast ion backgrounds from the single mode analysis in the framework of the quasilinear theory. Previous studies shown in Ref.~\onlinecite{Varenna} are here extended by including growth rates and frequencies obtained with the fast-ion numerical distribution functions. To resolve the fine velocity-space structure of the numerical backgrounds, $68$ points have been used for both the $v_{\shortparallel}$ and the $\mu$ GENE grids with simulations box sizes of respectively $(9, 3)$ in normalized units. For the analytical backgrounds, instead, $32$ points and $48$ equidistant symmetric grid points have been used for the $v_{\shortparallel}$ and $\mu$ GENE grids.
A first linear analysis is performed on the NBI fast Deuterium. In Fig.~\ref{fig:nbi} the GENE growth rates and frequencies are shown for different $k_{y} \rho_i$ values or equivalently for different toroidal mode numbers $n$. All the plasma species have been modelled with a local Maxwellian with the exception of the NBI fast Deuterium which, instead, has been modelled with the different analytical (slowing down) and numerical (NEMO/SPOT) distributions introduced in the previous paragraph. 
The growth rates and frequencies have been normalized to $c_{s} / a$ with $c_{s} = \left(T_{e}/m_{i}\right)^{1/2}$. A low sensitivity to the change of the fast Deuterium distribution function is observed. The velocity space anisotropies, well captured only from the numerical NEMO/SPOT distribution, do not significantly modify the linear results and only a relative difference of a few percent, i.e. $\leapp 10\%$, is observed. The slowing down distribution function can approximate better the numerical NEMO/SPOT results than the local Maxwellian. Furthermore, for this specific choice of fast Deuterium parameters, lower growth rates are found with the more realistic distributions.
A similar analysis can be performed for the ICRF-heated $^3$He. All the thermal plasma species have been modelled with a local Maxwellian while the NBI fast Deuterium either with a Maxwellian and a slowing down, which has been found to be the best analytical approximation to the NEMO/SPOT distribution.
\begin{figure*}
\begin{center}
\includegraphics[scale=0.25]{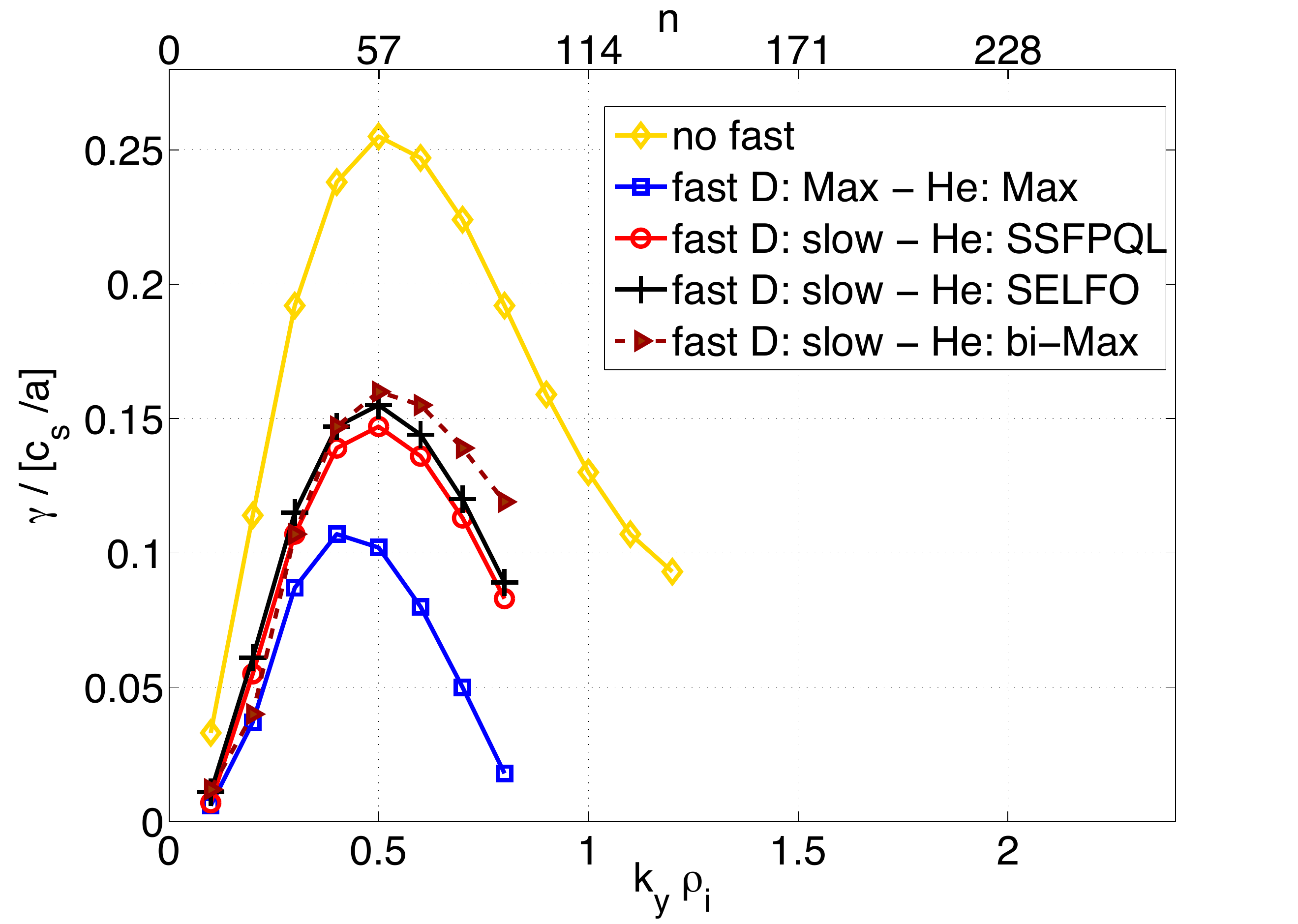}\includegraphics[scale=0.25]{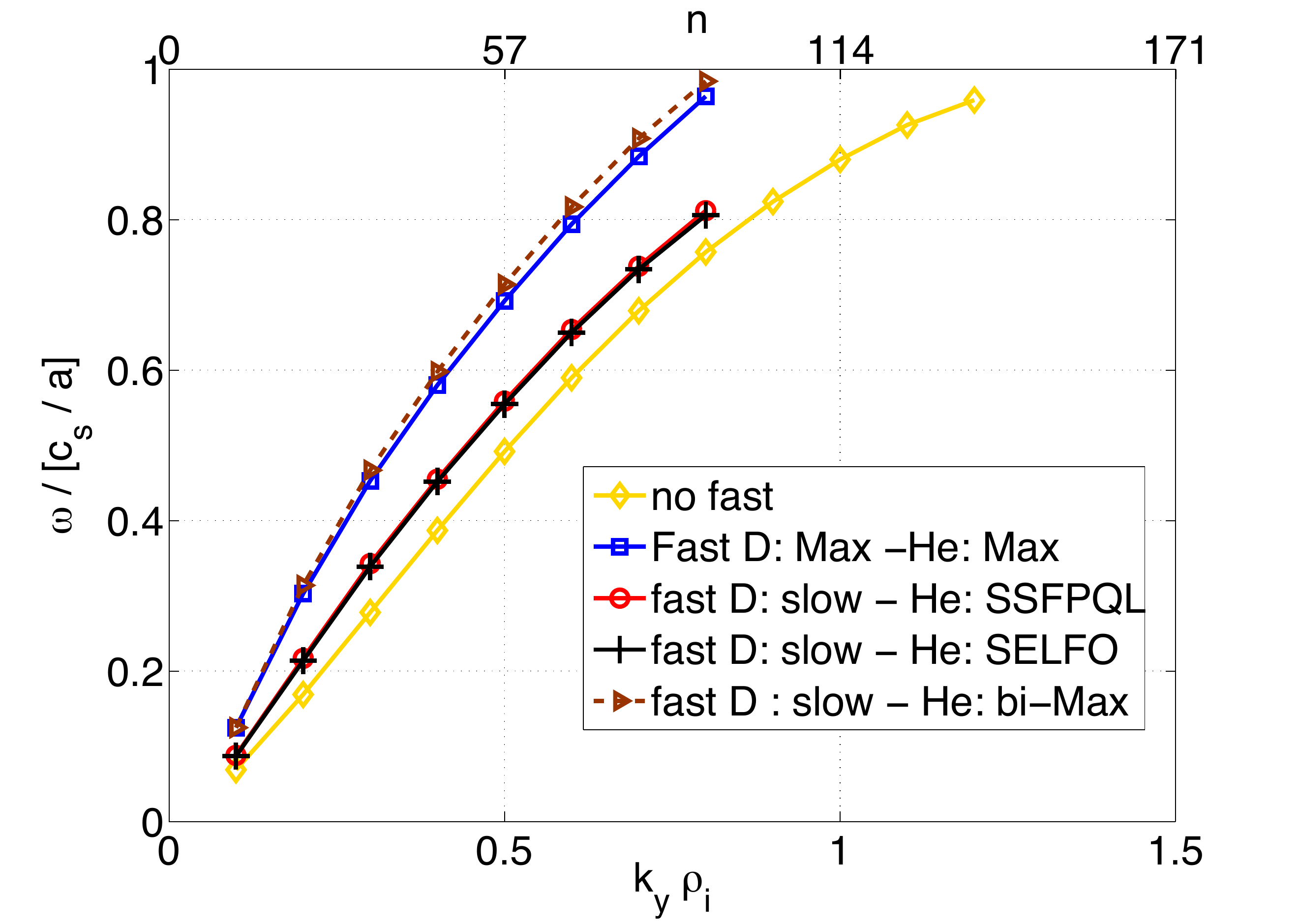}
\par\end{center}
\caption{GENE calculation of the linear growth rates (a) and frequency (b) for different $k_y \rho_i$ and toroidal mode numbers $n$ for different distribution functions for the fast $^3$He.}
\label{fig:icrh}
\end{figure*}
In Fig.~\ref{fig:icrh}, linear growth rates and frequencies are shown for different $^3$He backgrounds. In contrast to the previous results for fast deuterium, it is shown that the ICRH $^{3}$He has a significant impact on the linear ITG physics and differences of $\sim50\%$ are observed. A change in the background distribution and its radial derivative leads to a consequent change of the resonant ITG-fast ion stabilisation, which in Ref.~\onlinecite{Di_Siena} has been found to have significant effects on this discharge. For nominal parameters, the resonance ITG stabilising mechanism is predicted to be much more effective for the ICRF-heated $^{3}$He than for the NBI fast Deuterium, which might explain the lack of sensitivity of the fast Deuterium results on the different backgrounds. Moreover, with the more realistic $^{3}$He distribution functions a weakening of the still substantial fast ion stabilisation is observed. These results are consistent with experimental observations \cite{Citrin} and predict an overestimation of equivalent-Maxwellian fast-ion stabilisation for the nominal plasma parameters. According to quasilinear models an increase in the linear growth rates might lead to a relative increase of the nonlinear fluxes, greatly improving the agreement with experiments. Furthermore, an excellent agreement between the linear results obtained with TORIC/SSFPQL and SELFO/LION+FIDO is shown in Fig~\ref{fig:icrh}. The bi-Maxwellian has been found to be a good analytical approximation for the growth rate analysis to the numerical distributions in the low $k_y \rho_i\sim 0.1-0.3$ wave number range, where most of the transport typically originates in nonlinear ITG simulations.

\subsection{Turbulence analysis} \label{nonlinear}

\begin{figure*}
\begin{center}
\includegraphics[scale=0.22]{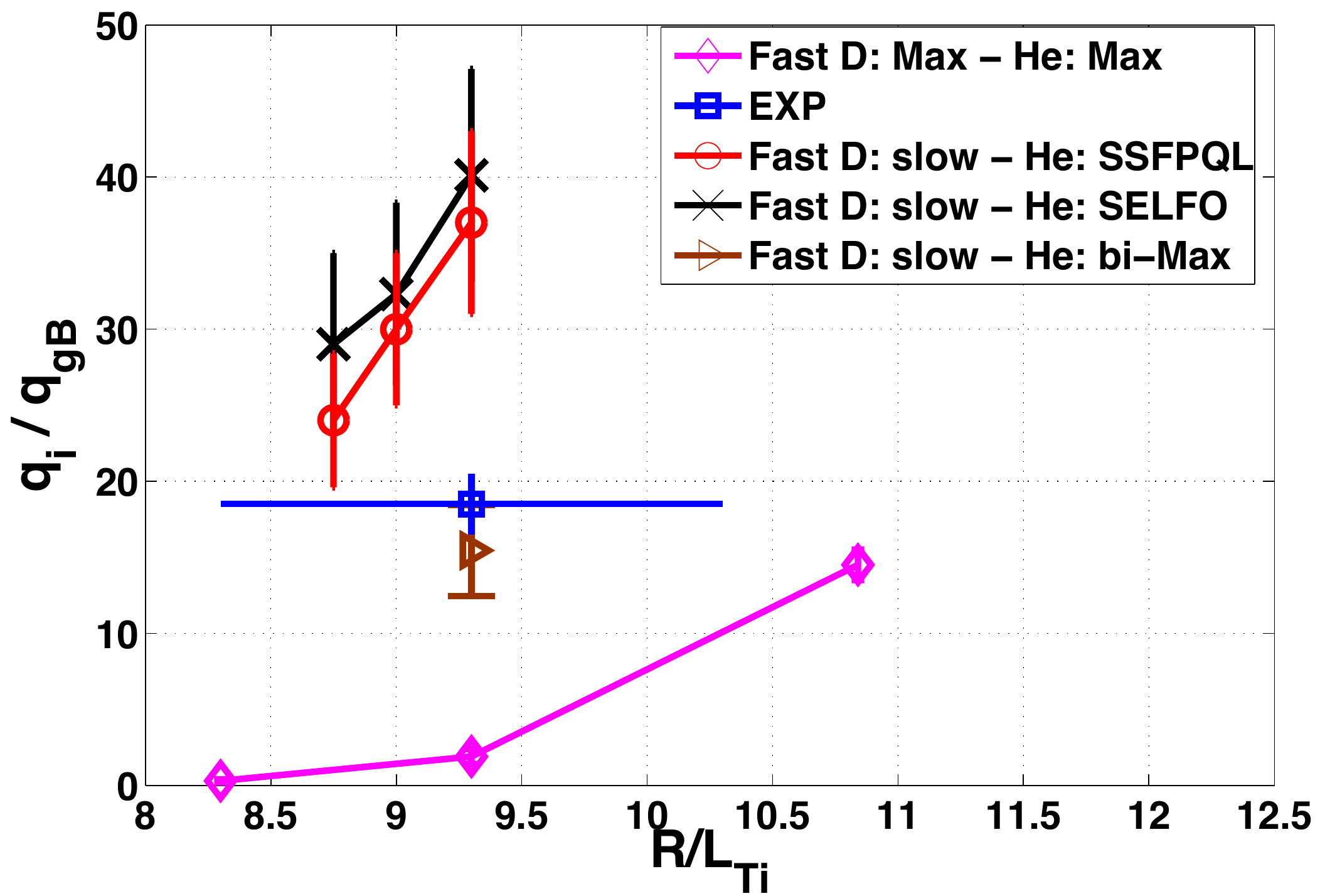}\includegraphics[scale=0.22]{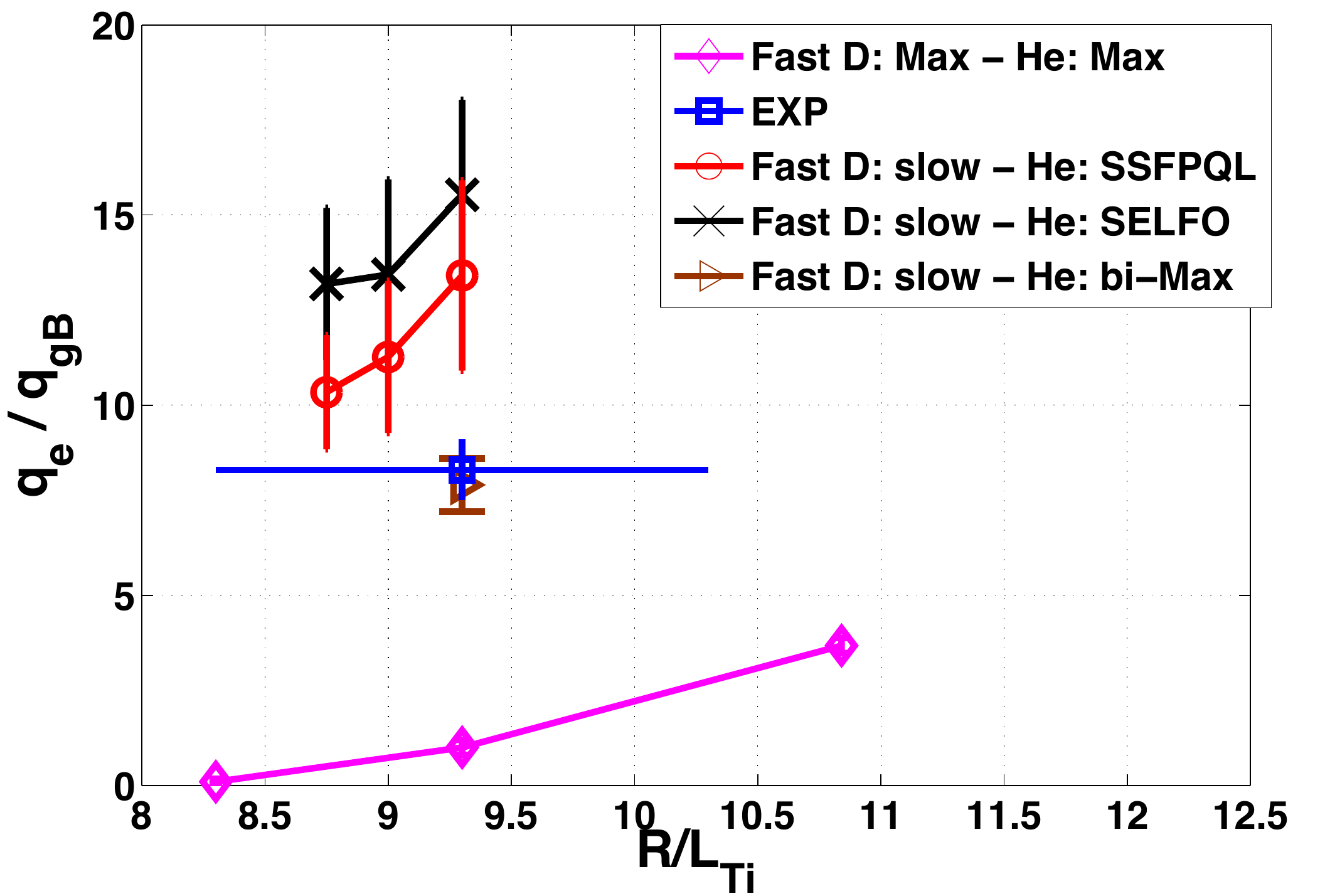}\includegraphics[scale=0.22]{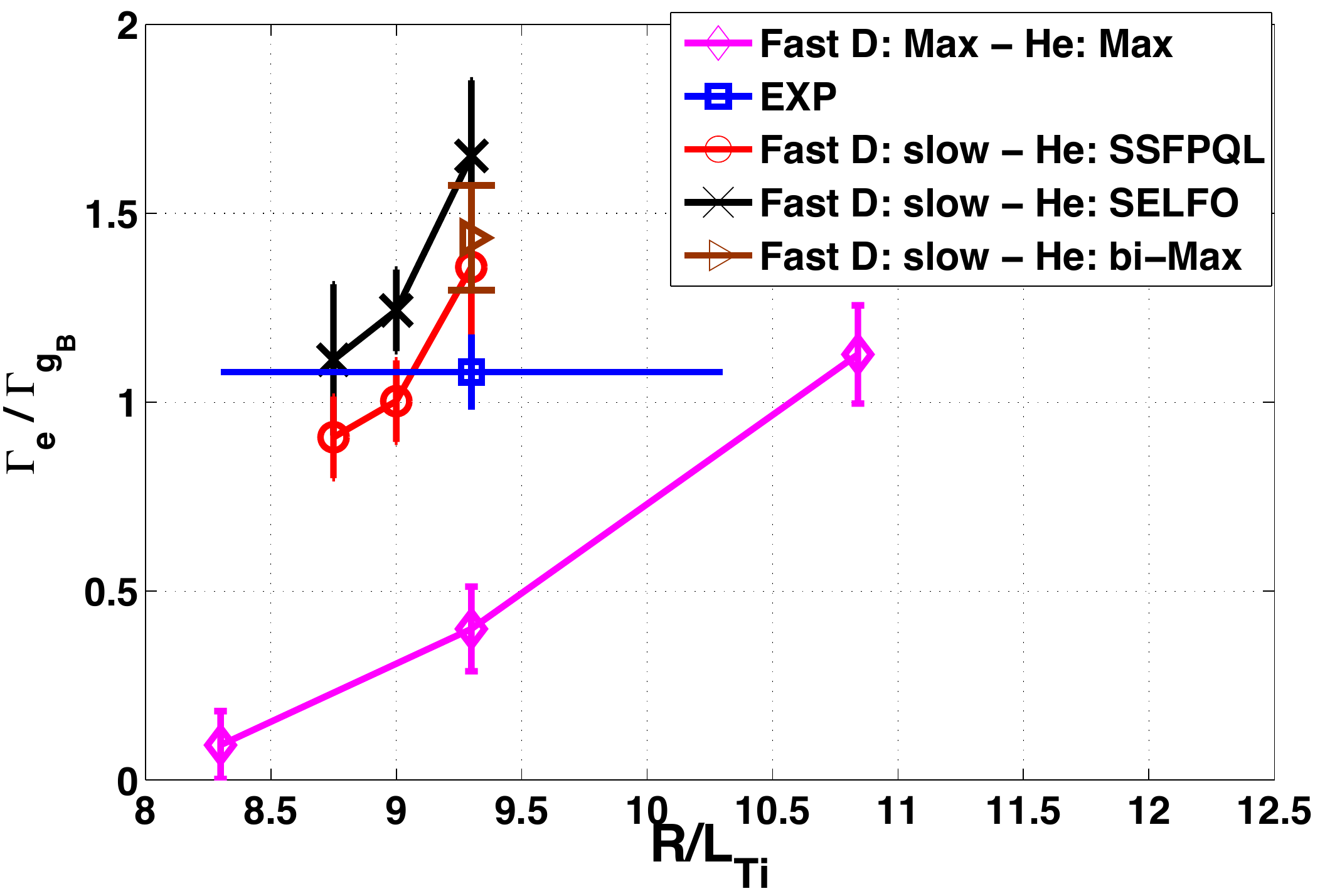}
\par\end{center}
\caption{Time-averaged nonlinear (a) main ion, (b) electron heat flux and (c) electron particle flux in GyroBohm units for different main ion temperature gradients and fast ion distribution functions.}
\label{fig:nonlinear_fluxes}
\end{figure*}

\begin{figure*}
\begin{center}
\includegraphics[scale=0.22]{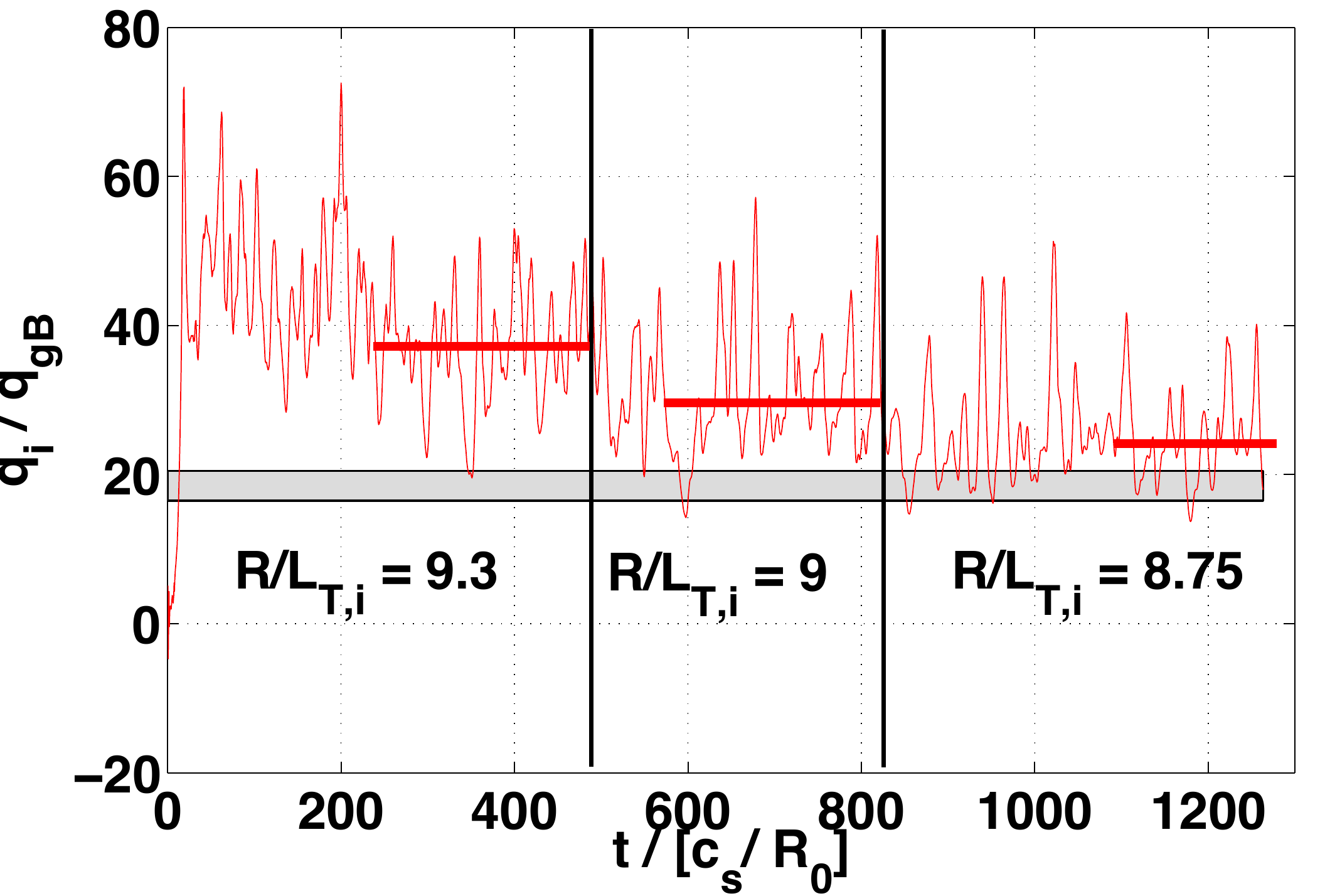}\includegraphics[scale=0.22]{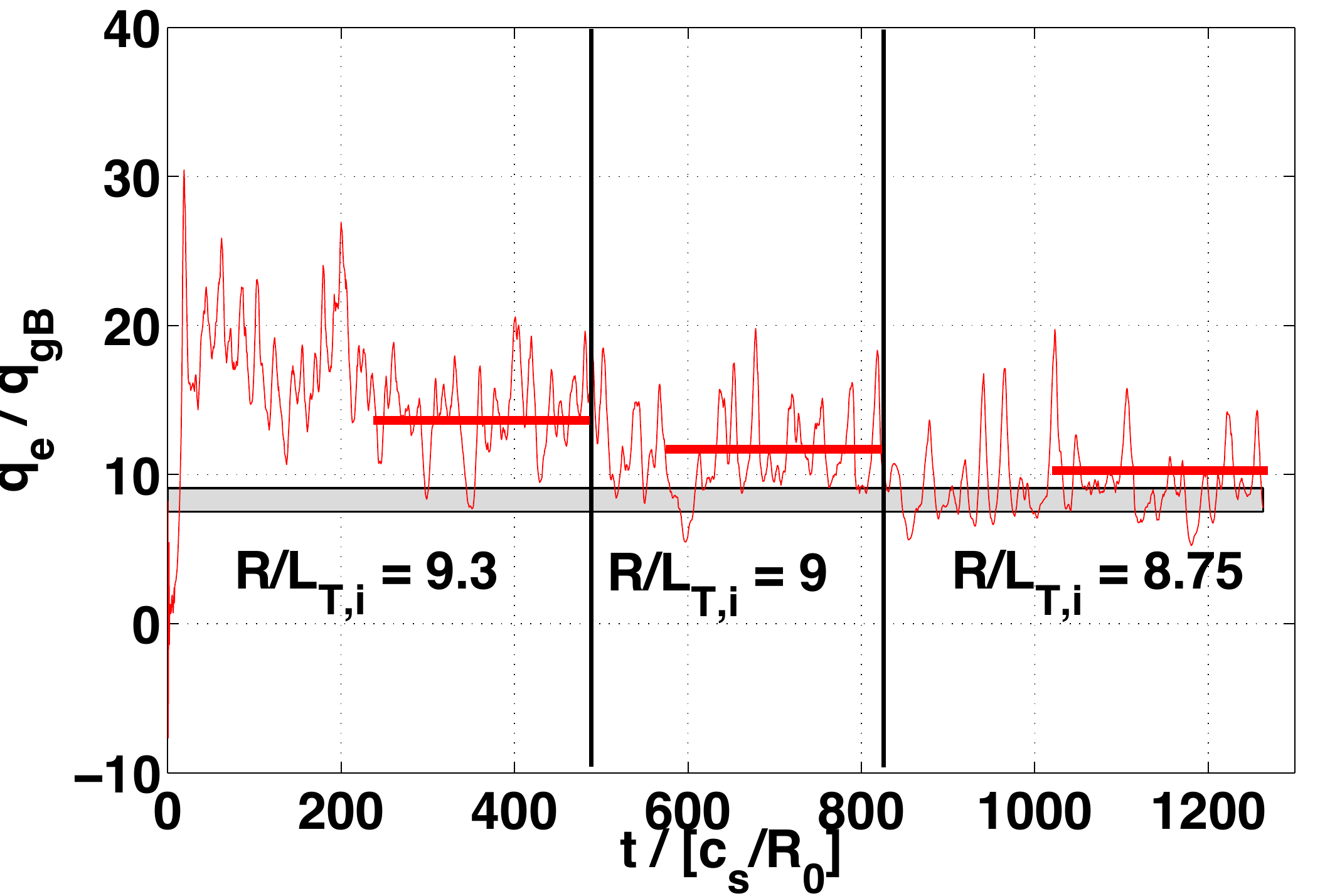}\includegraphics[scale=0.22]{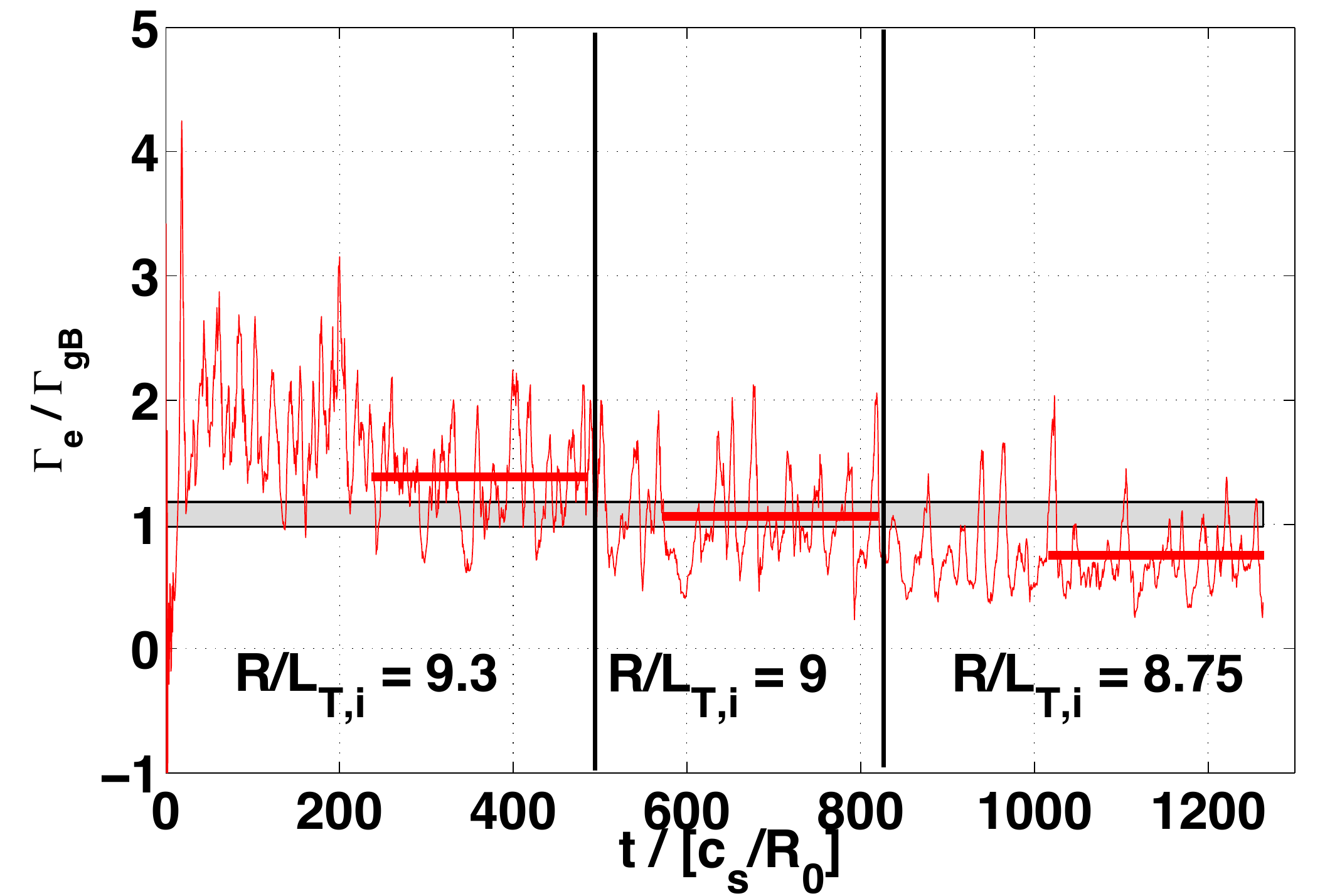}
\par\end{center}
\caption{Time trace of the nonlinear (a) main ion, (b) electron heat flux and (c) electron particle flux in GyroBohm units for different main ion temperature gradients for the case: slowing down fast deuterium and TORIC/SSFPQL fast helium. The gray area denotes the experimental value within error bars.}
\label{fig:time_fluxes}
\end{figure*}

\begin{figure}
\begin{center}
\includegraphics[scale=0.32]{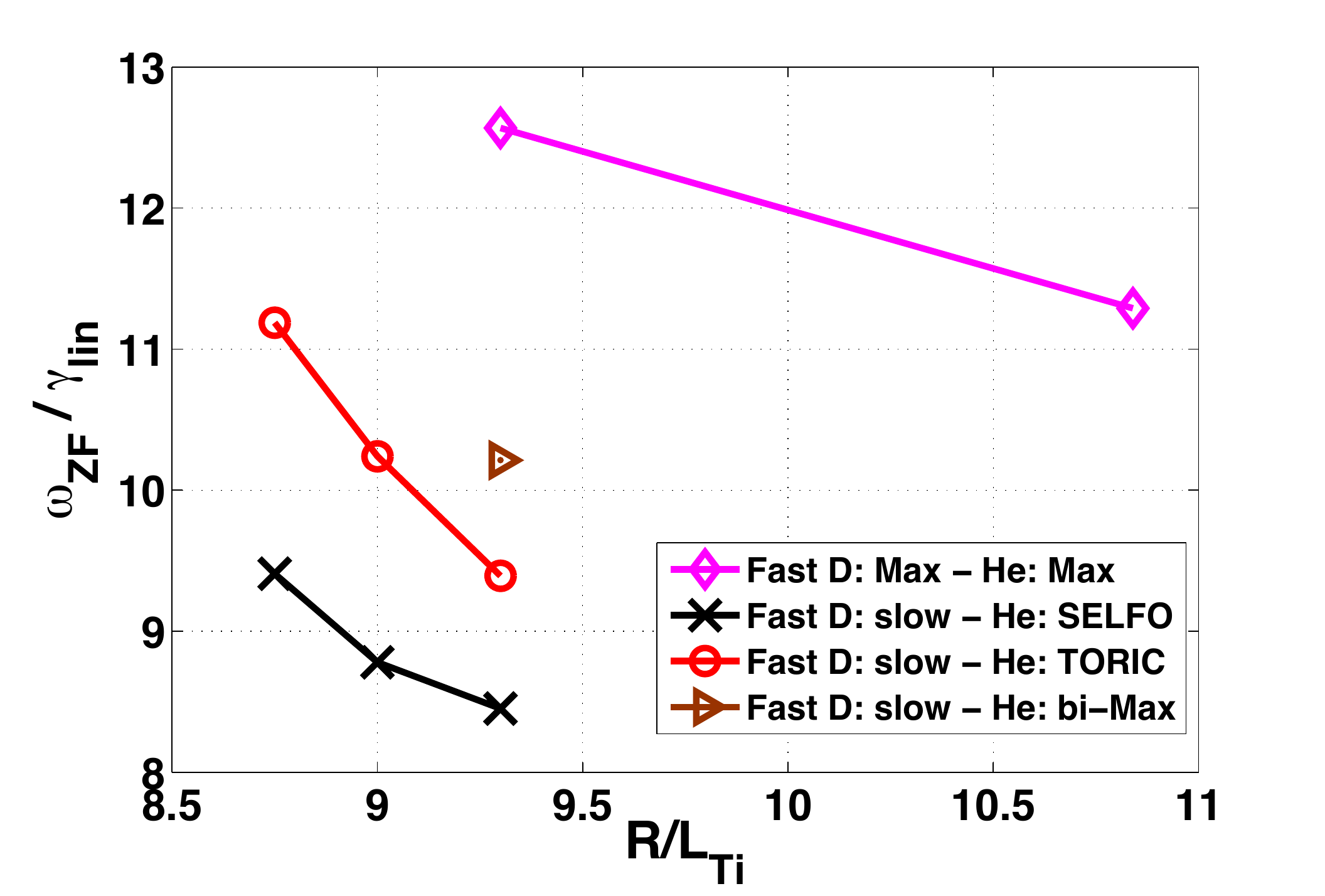}
\par\end{center}
\caption{Time-$k_x$ averaged $E \times B$ shearing rate normalised to the linear growth rate at the $k_y$ of the transport flux maximum for different main ion temperature gradients and for different fast ion distributions.}
\label{fig:shear}
\end{figure}

The impact of more realistic fast-ion distribution functions on the turbulent transport of the low-beta JET discharge 73224 is studied with GENE nonlinear simulations. The physical parameters are the same as in table ~\ref{table:parameters}. The radial box size is $175 \rho_i$ and the minimum $k_{y} \rho_i$ is set to $0.05$. We used $192$ grid points in radial direction, $48$ modes in the binormal direction and $32$ points along the field line. As for the linear simulations, a high velocity space resolution is required to resolve the fine velocity structure of the non-Maxwellian distribution functions. In velocity space, $68$ points and $68$ equidistant symmetric grid points have been used for the numerical distributions and $48$, $32$ for the analytical backgrounds for resolving respectively the $\mu$ and the $v_{\shortparallel}$ space with a $(\mu, v_{\shortparallel})$ box size of respectively $(9, 3)$ in normalized units. The first nonlinear analysis presented in this paper concerns the study of the main ion and electron fluxes. In a previous publication \cite{Citrin}, it has been shown that a much better agreement between the numerical and the experimental values - extracted from CRONOS \cite{CRONOS} interpretative simulations - was achievable only by including equivalent Maxwellian fast ions in the numerical simulations. However, the experimental fluxes were matched only by an increase of the main ion temperature gradient of $\sim20\%$ and an overestimation of the fast particle stabilizing effects was observed at the nominal plasma parameters.
In Fig.~\ref{fig:nonlinear_fluxes} a comparison between the nonlinear results obtained with the more realistic fast ion distribution functions is shown for values of the main ion temperature gradients inside the experimental error bars. In order to keep the same notation as in Ref.~\onlinecite{Mantica_PRL2009,Mantica_PRL2011,Citrin_PRL2013}, the particle and heat fluxes are normalised, respectively, to $\Gamma_{gB} = v_{th,i}\rho_i^2n_i/R_0^2$ and $Q_{gB} = v_{th,i}\rho_i^2n_iT_i/R_0^2$. Furthermore, the NBI fast deuterium has been modelled either with a local Maxwellian or with a slowing down distribution function. The NEMO/SPOT numerical distribution was numerically challenging in the full nonlinear GENE turbulence simulations. An increased number of markers is most likely required in NEMO/SPOT simulations in order to obtain a smoother numerical distribution compared to the coarse function with 4191 test particles used in this paper. However, as shown in the previous paragraph, no significant difference is expected by employing the numerical distribution function for the NBI fast deuterium.
The values of the fluxes are computed as a time-average over the saturated state of the simulations. In Fig.~\ref{fig:time_fluxes} the time trace of the main ion and electron fluxes obtained with slowing down NBI fast deuterium and numerical TORIC/SSFPQL helium is shown with the correspondent average value used for Fig.~\ref{fig:nonlinear_fluxes}. A significantly better agreement between numerical and experimental results is achieved with the more realistic distribution functions for the fast ion population. The experimental results are well reproduced by GENE simulations inside the temperature gradient error bars with both analytical (slowing down, bi-Maxwellian) and numerical (SSFPQL/TORIC-SELFO) distribution functions. In line with the linear results, a corresponding "weakening" of the (still significant) fast ion stabilisation is observed and, for the range of parameters here exploited, the bi-Maxwellian distribution is confirmed to be a good first order approximation to the more realistic backgrounds. Furthermore, a good agreement between GENE simulations based on TORIC and SELFO is here confirmed by the nonlinear results.
The impact of the different fast ion distribution functions on the nonlinear transport levels can be further investigated through the study of the zonal flow structure. It has been shown in several publications \cite{zonal1, zonal2, zonal3} that zonal flows - as major nonlinear saturation mechanisms - can play a significant role in the reduction of turbulent fluxes. In gyrokinetic simulations, zonal flow activity is often measured through the $E \times B$ shearing rate defined as follows
\begin{equation}
\omega_{ZF} = \frac{d^2 \phi_{zon}}{d^2 x}.
\label{eq:exb}
\end{equation}
Here, $\phi_{zon}$ is the zonal component of the electrostatic potential. In Fig.~\ref{fig:shear}, the ratio between $\omega_{ZF}$, averaged over all the $k_x$ mode components, and the linear growth rate at the $k_y$ of the transport flux maximum is shown for different values of the main ion temperature gradients and for the different fast ion distribution functions used in the nonlinear analysis of Fig.~\ref{fig:nonlinear_fluxes}. A qualitative though correlation between $<\omega_{ZF}>_{k_x} / \gamma_{lin}$ and the turbulent flux levels is observed in Fig.~\ref{fig:shear}. The zonal flow activity increases with a decrease of the main ion temperature gradients and lower fluxes are observed in GENE numerical simulations. These results suggest that the zonal flows are also affected by the more realistic fast-ion distribution functions and they are overestimated in the case of equivalent Maxwellian distributed fast ions. A more quantitative analysis will be done in future.

\section{Conclusions} \label{concl}

In the present paper the $\delta f$ gyrokinetic Vlasov-Maxwell coupled equations are re-derived for a completely arbitrary background distribution function in the full electromagnetic case. As a meaningful example for a possible application, a previous study on a particular low beta JET plasma with significant fast ion stabilisation is revised with more realistic distribution functions for the fast ion population compared to the results obtained with equivalent Maxwellian background distributions. The bulk plasma is composed by Deuterium, electron and Carbon impurities, while the fast particles are NBI fast deuterium and ICRH accelerated $^3$He. Electromagnetic effects, collisions and experimental geometry are taken into account in the simulations. In the linear analysis it is found that with the more realistic distribution functions the fast ion stabilisation still holds, even if it is weakened. This is in line with the previous nonlinear findings where gradients higher than the nominal ones had to be employed in order to match the experimental heat fluxes in the presence of fast particles \cite{Citrin_PRL2013}. The impact of the different non-Maxwellian backgrounds is studied separately on each fast ion species and a lack of sensitivity to the NBI fast ion distribution is observed. Generally, the choice of the $^3$He background distribution - particular, its anisotropies and asymmetries - has a stronger impact on the linear results than the fast deuterium backgrounds. As discussed in this paper, a change in the background distribution affects the resonant ITG-fast ion stabilisation which, for this JET discharge, is particularly strong only for the fast helium population, as proved in Ref.~\onlinecite{Di_Siena}. These linear results are confirmed by the GENE nonlinear turbulence simulations. An improved agreement between the experimental and numerical results is achieved for the main ion and electron fluxes at the nominal plasma parameters when more realistic fast-ion distribution functions are employed. Additionally, for the range of parameters considered here, the bi-Maxwellian and the slowing down distributions are shown to be good first order approximations to respectively the fast helium and deuterium numerical backgrounds. A good agreement between nonlinear GENE results obtained using TORIC/SSFPQL and SELFO/LION+FIDO distribution functions is here confirmed. First results suggest that the choice of the background distribution function has also an impacts on the level of zonal-flow activity.

\section*{Acknowledgement}
The simulations presented in this work were performed using the HYDRA cluster at the Rechenzentrum Garching (RZG), Germany. Furthermore, we acknowledge the CINECA award under the ISCRA initiative, for the availability of high performance computing resources and support. This work has been carried out within the framework of the EUROfusion Consortium and has received funding from the Euratom research and training programme 2014-2018 under grant agreement No 633053. The views and opinions expressed herein do not necessarily reflect those of the European Commission. 
The author would like to thank F. Jenko, Ph. Lauber and M. Romanelli for all the stimulating discussions, useful suggestions and comments.

\end{document}